\documentclass[groupedaddress,aps,prd,preprintnumbers,onecolumn,11pt,fleqn,nofootinbib,nobibnote,eqsecnum,floatfix,showpacs]{revtex4-1}

\usepackage{graphicx}
\usepackage{bm,amsmath,amssymb}
\usepackage[mathscr]{eucal}

\long\def\comment#1{ }
\newcommand{\eqn}[1]{Eq.~\eqref{#1}}
\newcommand{\beq}{\begin{equation}}
\newcommand{\eeq}{\end{equation}}
\newcommand{\nn}{\nonumber\\}
\newcommand{\dif}{{\rm d}}
\newcommand{\rmd}{{\rm d}}

\newcommand{\rmi}{{\rm i}}
\newcommand{\rmtr}{{\rm tr}}

\newcommand{\del}{\partial}
\newcommand{\lan}{\langle}
\newcommand{\ran}{\rangle}
\newcommand{\blan}{\big\langle}
\newcommand{\bran}{\big\rangle}
\newcommand{\order}[1]{\mcal{O}{(#1)}}
\newcommand{\mcal}{\mathcal}
\newcommand{\wt}{\widetilde}
\newcommand{\bmk}{\bm{k}}
\newcommand{\bmx}{\bm{x}}
\newcommand{\bmy}{\bm{y}}
\newcommand{\bmu}{\bm{u}}
\newcommand{\bmv}{\bm{v}}
\newcommand{\bmz}{\bm{z}}

\newcommand{\atpi}{\frac{\bar{\alpha}}{2 \pi}}
\newcommand{\abar}{\bar{\alpha}}

\begin{document}

\preprint{arXiv:1212:1656}

\title{Testing the Gaussian Approximation to the JIMWLK Equation}
\author{M.~Alvioli}
 \email{alvioli@pg.infn.it}
 \altaffiliation{Current affiliation: {\it CNR-IRPI, via Madonna Alta 126, I-06128 Perugia (PG), Italy}}
 \affiliation{European Centre for Theoretical Studies in Nuclear Physics and Related Areas (ECT*), Strada delle Tabarelle 286, I-38123 Villazzano (TN), Italy}
\author{G.~Soyez}
 \email{gregory.soyez@cea.fr}
 \affiliation{Institut de Physique Th\'{e}orique, CEA/Saclay, F-91191 Gif-sur-Yvette, France}
\author{D.N.~Triantafyllopoulos}
 \email{trianta@ectstar.eu}
 \affiliation{European Centre for Theoretical Studies in Nuclear Physics and Related Areas (ECT*)\\ and Fondazione Bruno Kessler, Strada delle Tabarelle 286, I-38123 Villazzano (TN), Italy}

\date{\today}

\begin{abstract}
In processes involving small-$x$ partons, like in deep inelastic scattering and in 
hadronic collisions at high energy, the final state can be expressed in terms of 
correlators of Wilson lines. We study such high-point correlators evolving 
according to the JIMWLK equation and we confirm the results of previous numerical 
and analytic work, by using an independent method, that the solution to the JIMWLK 
equation can be very well approximated by an appropriate Gaussian wavefunction. We explore both fixed and running coupling evolution, where in the latter the scale is set according to various prescriptions. As 
a byproduct, we also numerically confirm to high accuracy the validity of the law 
governing the behavior of the $S$-matrix close to the unitarity limit, the 
Levin-Tuchin formula. We furthermore outline how to calculate correlators with open color indices.
\end{abstract}

%\keywords{Color Glass Condensate, parton saturation, JIMWLK equation, particle production}

\maketitle

%\tableofcontents

\section{\label{sec:intro}Introduction and motivation}

In hadronic collisions at ultra-relativistic energies the final state is quite
involved in terms of the type, the number and the distribution of the produced 
particles. The extraction of the dominant physical mechanisms in such processes 
is a demanding task and in order to achieve the best possible understanding it 
is necessary to study many observables in wide kinematic regimes. For 
instance, significant attention has been given to collisions between light and heavy 
hadrons, like deuteron-gold at RHIC and the forthcoming proton-lead at the 
LHC and in both cases two of the most representative observables are related to 
single and double inclusive particle production. Considering the
single inclusive particle production 
in the deuteron fragmentation region it has been observed, already a few 
years ago at RHIC \cite{Arsene:2004ux,Adams:2006uz}, a suppression compared to 
the cross section obtained by taking the nucleus as an incoherent superposition 
of $A^{1/3}$ nucleons. More recently, and regarding the double
inclusive particle production, 
it has been seen an increasing suppression of the azimuthal correlation of the 
two hadrons when their transverse momenta are a few GeV 
\cite{Braidot:2010ig,Adare:2011sc}, as we move towards the deuteron 
fragmentation region. 

This is precisely the kinematic regime which encourages the search for parton
saturation in the wave-function of the heavy hadron, the large nucleus. 
Natural qualitative and quantitative descriptions of the RHIC data and predictions
for the LHC upcoming ones based on such a physical mechanism, along with the 
corresponding formulations, already exist for the nuclear modification factor 
$R_{pA}$ \cite{Kharzeev:2003wz,Albacete:2003iq,Blaizot:2004wu,Iancu:2004bx,
Albacete:2010bs,Altinoluk:2011qy,JalilianMarian:2011dt,
Chirilli:2011km,Tribedy:2011aa,Mueller:2012bn,Albacete:2012xq,Rezaeian:2012ye}, 
which is related to the single inclusive cross section. Similarly, the di-hadron 
azimuthal correlations at RHIC offer a unique environment to test parton saturation \cite{JalilianMarian:2004da,Marquet:2007vb} 
and in fact the corresponding data have been understood in that context 
\cite{Tuchin:2009nf,Albacete:2010pg,Stasto:2011ru,Lappi:2012nh}.

\begin{figure}
 \begin{center}
 \begin{minipage}[b]{0.325\textwidth}
 \begin{center}
 \includegraphics[scale=0.55]{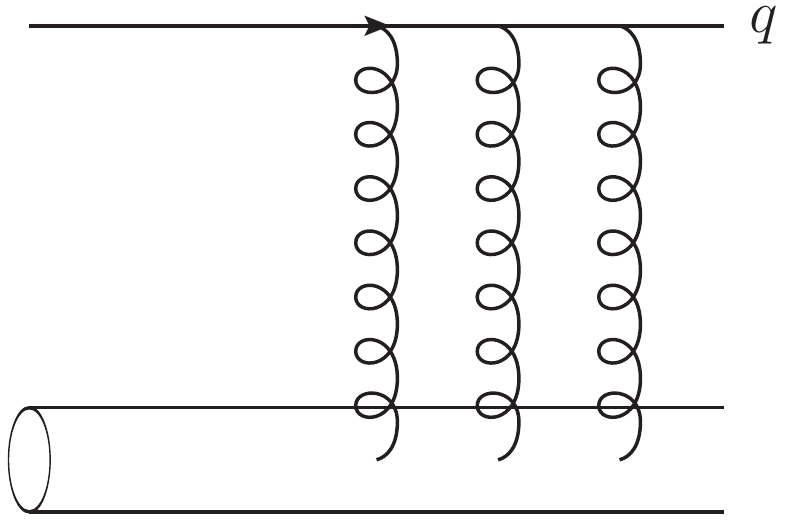}\\{\footnotesize (a)}
 \end{center}
 \end{minipage}
 \begin{minipage}[b]{0.325\textwidth}
 \begin{center}
 \includegraphics[scale=0.55]{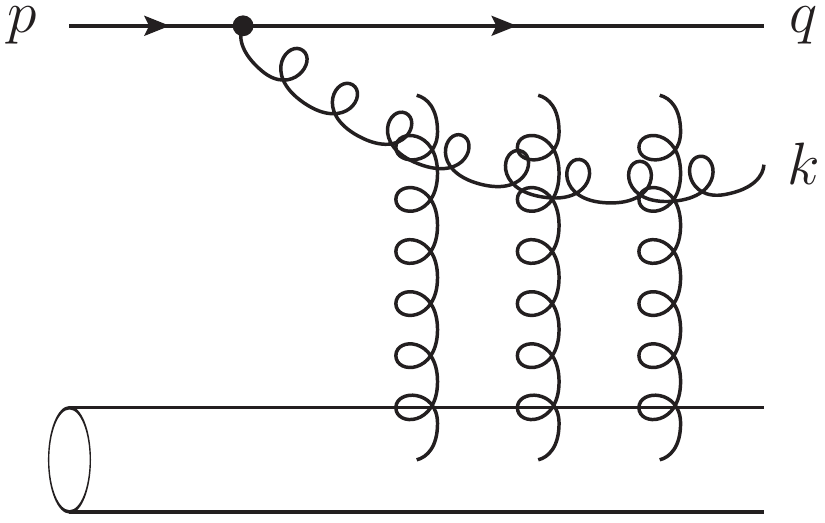}\\{\footnotesize (b)}
 \end{center}
 \end{minipage}
 \begin{minipage}[b]{0.325\textwidth}
 \begin{center}
 \includegraphics[scale=0.55]{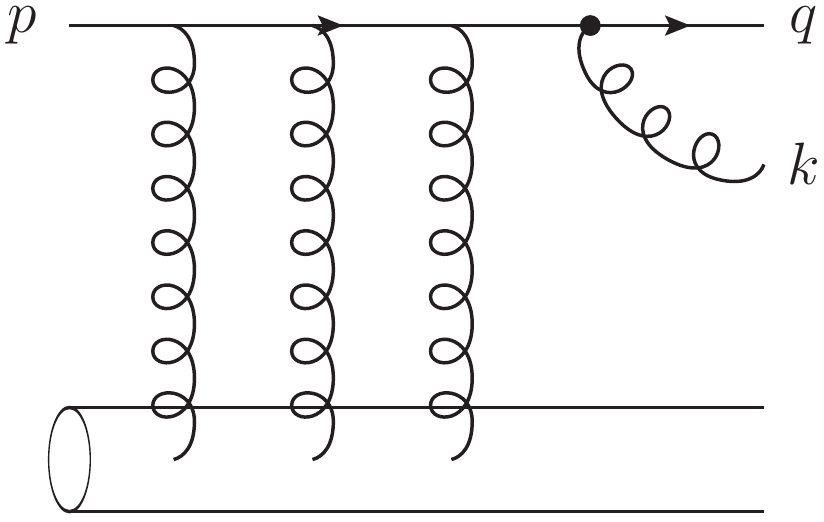}\\{\footnotesize (c)}
 \end{center}
 \end{minipage}
 \caption{(a) Quark production and (b), (c) quark-gluon production in $pA$ collisions.\label{fig:jet}}
 \end{center}
 \end{figure}

Thus, one is particularly interested in $hA \to h'X$ and $hA \to h_1 h_2 X$,
with $h$ a projectile hadron whose wavefunction is not saturated, like a proton at 
not too high energy so that its small-$x$ evolution can be neglected, and $A$ a 
target who can be dense, like an ultra-relativistic 
heavy nucleus. Let us again look at single particle production first for which the 
corresponding diagram at the partonic level, say for quark production, is shown in 
Fig.~\ref{fig:jet}.(a). A large-$x$ quark from the projectile interacts via 
multiple gluon exchanges with the small-$x$ components of the target and then it is
measured in the region which is forward in (pseudo)rapidity. The target is viewed 
as a potentially large color field $\mcal{A}^{\mu}_a$, the Color Glass Condensate 
(CGC) (see e.g.~\cite{Iancu:2002xk}) and the interaction of a parton with transverse 
position $\bmx$ with such a field is described by a Wilson line along its 
trajectory. Taking the modulus squared of 
the diagram \ref{fig:jet}.(a) in coordinate space, averaging over initial colors and 
summing over final ones, we find the cross section $qA \to q X$ 
to be given by the Fourier transform of a color dipole $\hat{S}$, that is, a trace 
of two Wilson lines in the fundamental representation, which is an overall colorless 
object.  

The above discussion naturally extends to the case of double particle production
when both particles are detected at the same rapidities. Without any loss 
of generality, les us focus on $qg$ production\footnote{Here and in
  the previous paragraph, we discuss only representative cases of
  single and double inclusive particle production which are taken to
  be quark and quark-gluon production respectively. Other
  possibilities like gluon and gluon-pair production have been also
  studied and understood
  \cite{Kovchegov:2001sc,JalilianMarian:2004da,Kovner:2006wr,Gelis:2008sz,Dusling:2009ni,Kovchegov:2012nd}.} at the partonic level with the 
respective diagrams shown in Fig.~\ref{fig:jet}.(b) and (c). The large-$x$ 
projectile quark splits into a quark-gluon pair, and this splitting can occur 
either before or after the interaction with the target. Now one arrives at an 
inclusive $qA \to qgX$ cross section \cite{Marquet:2007vb} which is 
given by a Fourier transform of various terms involving two, four or six Wilson 
lines. Such a term with the maximal number of Wilson lines is $\hat{S}\hat{Q}$, that is, a color dipole times  a color quadrupole, where the latter is a trace of four Wilson lines, and in fact it 
is not hard to understand the counting of Wilson lines. For instance, diagram (b) 
involves one in the fundamental and one in the adjoint representation and the 
latter can be expressed in terms of two fundamental ones. When multiplied with its 
complex conjugate it gives rise to the aforementioned $\hat{S}\hat{Q}$ term which 
involves six Wilson lines (plus a $1/N_c^2$ correction). Clearly, one realizes that 
the production of more and more particles at forward rapidities will involve more 
and more Wilson lines. Interestingly enough, it was recently shown that
in the large-$N_c$ limit any forward multi-particle production cross section can be 
expressed only in terms of dipoles and quadrupoles \cite{Dominguez:2012ad}.

Thus, in general, one needs to calculate correlators of the form 
$\lan \hat{O}\ran_Y$, where  $\hat{O}$ is constructed from such multipoles. 
Obviously the Wilson lines depend on the target field and the average has to be 
taken with the target probability distribution $W_{Y}[\mcal{A}]$, which simply 
gives the probability to find a given configuration in the target wavefunction.
The rapidity $Y$ is determined by the kinematics of the process under 
consideration; for example, for $qg$ production at forward rapidities one has  
$x=\exp(-Y) = [|\bmk|\exp(-|y_k|) + |\bm{q}|\exp(-|y_q|)]/\sqrt{s}$, 
with $\bm{q}$ and $y_q$ the transverse momentum and (pseudo)rapidity 
of the produced quark, $\bm{k}$ and $y_k$ those of the gluon and 
$\sqrt{s}$ the center of mass energy. At moderately 
small values of $x$ one typically invokes the McLerran-Venugopalan (MV) model 
\cite{McLerran:1993ni,McLerran:1993ka}, 
which is equivalent to a Gaussian wavefunction $W_Y[\alpha]$ and thus allows for explicit 
calculation of high-point correlators 
\cite{Blaizot:2004wv,JalilianMarian:2004da,Dominguez:2011wm}.

As we move towards higher values of $y_q$ and $y_k$, we probe components of the target with smaller-$x$ and it becomes necessary to 
take into account the evolution in $Y$ of $W_{Y}[\mcal{A}]$. Even though this 
is a classical probability distribution its evolution is quantum and satisfies
the JIMWLK equation 
\cite{JalilianMarian:1997jx,JalilianMarian:1997gr,Weigert:2000gi,
Iancu:2000hn,Iancu:2001ad} which arises from the resummation of the dominant
terms in $\ln(1/x)$ in the presence of a potentially strong color field. Using the Langevin form of the JIMWLK equation \cite{Blaizot:2002xy}, a critical 
work on its numerical solution was recently performed, where the very good accuracy 
of a particular Gaussian approximation, the extension to arbitrary $Y$ of the MV 
model, was observed for some configurations of high-point correlators \cite{Dumitru:2011vk}. Immediately after it was analytically
understood why the approximation scheme based on a Gaussian wavefunction provides a quasi-exact solution to the 
JIMWLK equation \cite{Iancu:2011ns,Iancu:2011nj}. 

Our effort, which is mostly numerical, may be considered as complementary to the 
work in \cite{Dumitru:2011vk} to some extent, but it is completely independent 
for a number of reasons. (i) The method here is different; instead of studying 
JIMWLK as a Langevin equation on a lattice, we directly analyze evolution equations 
for given configurations. But even if one starts from a simple comfiguration, evolution always creates more general ones (cf.~Figs.~\ref{fig:configs1} and \ref{fig:configs2}) and eventually our tests probe a much wider sample. (ii) We shall explore a larger kinematical space and as a 
byproduct we shall be able to give a numerical verification of the Levin-Tuchin 
formula \cite{Levin:1999mw} (see also 
\cite{Mueller:1996te,Iancu:2001md,Mueller:2002pi,Iancu:2003zr}). (iii) We will study 
both fixed and running coupling evolution, where the scheme in the latter case can 
be chosen at our will. (iv) We can consider any value for the number of colors 
$N_c$, even though we shall mainly work in the multicolor limit. 

The paper is mostly devoted to technical aspects concerning the validity of a 
certain approximation scheme and its structure is the following: 
In Sect.~\ref{sec:jimwlk} we review the JIMWLK equation and how it determines 
the evolution of multi-parton correlations. In this Section we also introduce most 
of our definitions along with the corresponding notation. In Sect.~\ref{sec:bk} we 
give an accurate numerical solution to the Balitsky-Kovchegov (BK) equation, that 
is, the large-$N_c$ equation for the dipole 
\cite{Balitsky:1995ub,Kovchegov:1999yj}. In particular we focus on the regime where 
saturation has been reached, a necessary step for our purposes, and we verify for 
the first time the validity of the Levin-Tuchin formula. Moreover, from the form of 
the solution at saturation, we also confirm the two most dominant terms in the 
asymptotic expansion of the saturation momentum $Q_s$ for fixed coupling evolution. 
In Sect.~\ref{sec:gauss} we briefly reflect on the Gaussian approximation and why 
it is expected to work accurately \cite{Iancu:2011ns,Iancu:2011nj}. 
In Sect.~\ref{sec:mfa} we study numerically a Mean Field Equation arising from the 
Gaussian approximation to the JIMWLK Hamiltonian and show that it is in agreement 
with the BK equation to logarithmic accuracy at saturation as expected. In Sect.~\ref{sec:quad} we examine
two quadrupole configurations and show, for both fixed and running coupling evolution (the latter in the Balitsky, smallest dipole and daughter dipole prescriptions), that the accuracy is not restricted to 
be logarithmic, giving further evidence that the Gaussian approximation is a 
``quasi-exact'' solution. We point out a potential deficiency of a ``too simple'' Gaussian approximation (in particular the extrapolation to arbitrary $Y$ of the MV model) in the running coupling case and at saturation, but it seems not to be crucial for practical purposes.  In Sect.~\ref{sec:open} we outline how one can 
calculate correlators with open color indices and finally in Sect.~\ref{sec:conclusion} we conclude.

\section{\label{sec:jimwlk}The JIMWLK equation and multi-parton correlations}

The Color Glass Condensate (CGC) is a modern effective theory for the small-$x$ 
components of the wavefunction of an ultra-relativistic hadron. It relies on the 
idea that gluons which carry a small fraction $x$ of the hadron's longitudinal 
momentum can be described as a random distribution of classical color fields 
generated by sources with larger momentum fractions. As a result of the high energy 
kinematics, the distribution of the color sources is frozen due to Lorentz time 
dilation and the color field, in a suitable gauge, has a single non-zero component. 
More precisely, and using the standard definitions for the light-cone coordinates 
$x^\mu=(x^+,x^-,\bmx)$, with $x^\pm=(t\pm x^3)/\sqrt{2}$ and
$\bmx=(x^1,x^2)$, this gauge field is 
$\mcal{A}^\mu_a(x) =\delta^{\mu+}\alpha_a(x^-,\bmx)$ if the hadron moves along the 
positive $x^3$-axis. The CGC weight function $W_Y[\alpha]$ gives the probability 
that the hadron be described by the configuration $\alpha$ and it is a functional 
probability distribution whose knowledge allows the determination of the 
correlations of the gauge field. The latter contain all the detailed information 
about the evolution of the hadron with increasing rapidity $Y\equiv \ln(1/x)$, from 
an initial value $Y_0=\ln(1/x_0)$ to the value $Y$ of interest. At high energy 
where $\alpha_s (Y-Y_0)\gtrsim 1$ and to leading order with respect to the large 
logarithm $\ln(x_0/x)$, this evolution obeys a renormalization group equation, the 
JIMWLK equation 
\cite{JalilianMarian:1997jx,JalilianMarian:1997gr,Weigert:2000gi,
Iancu:2000hn,Iancu:2001ad}. In a Hamiltonian form it reads
 \beq\label{jimwlk}
 \frac{\del}{\del Y}W_Y[\alpha]
 =  H W_Y[\alpha]\,,
 \eeq
where $H$ is the JIMWLK Hamiltonian and it is a second-order functional
differential operator whose most elegant and convenient for our purposes form was given is \cite{Hatta:2005as}
 \beq\label{H}
 H = -\frac{1}{16 \pi^3} \int_{\bmu\bmv\bmz}
 \mcal{M}_{\bmu\bmv\bmz}
 \left(1 + \wt{V}^{\dagger}_{\bmu} \wt{V}_{\bmv}^{\phantom{\dagger}}
 -\wt{V}^{\dagger}_{\bmu} \wt{V}_{\bmz}^{\phantom{\dagger}}
 -\wt{V}^{\dagger}_{\bmz} \wt{V}_{\bmv}^{\phantom{\dagger}}\right)^{ab}
 \frac{\delta}{\delta \alpha_{\bmu}^a}
 \frac{\delta}{\delta \alpha_{\bmv}^b}.
 \eeq
Here we have used the economical notation
$\int_{\bmu \dots} \equiv \int \dif^2 \bmu \dots$, 
defined the dipole kernel $\mcal{M}$ \cite{Mueller:1993rr}
 \beq
 \mcal{M}_{\bmu\bmv\bmz}\,
 \equiv\, 
 \frac{(\bm{u}-\bm{v})^2}{(\bm{u}-\bm{z})^2(\bm{z}-\bm{v})^2}
 \eeq
and introduced the Wilson lines
 \beq\wt{V}^{\dagger}_{\bmx}\,\equiv\,{\mbox P}\exp 
 \left[
 \rmi g\int \rmd x^-\alpha_a(x^-,\bmx)T^a \right], \label{Vadj}
 \eeq
where $T^a$ is in the adjoint representation and with P denoting path-ordering in 
$x^-$. The precise action of the functional derivatives appearing in \eqn{H} will 
be explained later on. The above form of the Hamiltonian is valid only when acting 
on gauge-invariant functionals of $\alpha_a$, like gauge-invariant products of 
Wilson lines. This will be the case for most of our analysis with exceptions to be 
discussed at the end in Sect.~\ref{sec:open}. Needless to say, in order to specify 
our problem completely, we need an initial condition for \eqn{jimwlk} at $Y=Y_0$. 
At least for a sufficiently large nucleus ($A \gg 1$, with $A$ the atomic mass 
number), this initial condition is typically provided by the McLerran-Venugopalan 
(MV) model \cite{McLerran:1993ni,McLerran:1993ka}.

Physical observables are represented by gauge invariant operators 
$\hat{\mcal{O}}[\alpha]$ constructed with the gauge color field $\alpha_a$. Their 
expectation value can be computed as a functional average with the CGC
weight function, that is
 \beq \label{average}
 \langle \hat{\mathcal O}\rangle_Y \equiv 
 \int{\mathcal D}\alpha \, {\mathcal O}[\alpha]\, W_Y[\alpha].
 \eeq
The above makes clear that, even though $W_Y[\alpha]$ is obtained by a quantum 
calculation, the averaging procedure is classical. Differentiating the above with 
respect to $Y$, using \eqn{jimwlk}, and finally  integrating twice by parts, we 
arrive at the evolution equation
 \beq\label{general}
 \frac{\del \lan \hat{\mcal{O}} \ran_Y}{\del Y}
 = \lan H \hat{\mcal{O}} \ran_Y\,
 \eeq
for the observable under consideration. This is not a functional equation anymore, 
but an integro-differential equation as we shall see in a while in specific 
examples. Still, this is not much easier to deal with since, due to the non-linear 
dependence of the Hamiltonian \eqref{H} on the field $\alpha_a$, \eqn{general} is 
in general not a closed equation, but just a member of an infinite hierarchy of 
coupled equations, the Balitsky equations \cite{Balitsky:1995ub,Balitsky:2001gj}.
The JILWLK equation \eqref{jimwlk} and the Balitsky hierarchy offer complementary 
views on the high-energy evolution. On the one hand, \eqn{jimwlk} describes the 
evolution of the target by the emission of an additional gluon with rapidity 
between $Y$ and $Y+\rmd Y$ in the background of the gauge color field $\alpha$ 
built by previous emissions, at rapidities smaller than $Y$. The Wilson lines in 
the Hamiltonian \eqref{H} correspond to the propagation of this new gluon in the 
background field and in the eikonal approximation. On the other hand, the Balitsky 
hierarchy focuses on the projectile evolution and in particular on the operator 
describing its scattering off the target. This scattering is again considered in 
the eikonal approximation and therefore the operator
$\hat{\mcal{O}}$ is naturally constructed with Wilson lines, where each one of them 
represents a parton in the projectile.

\begin{figure}
\begin{minipage}[b]{0.245\textwidth}
\begin{center}
\includegraphics[scale=0.5]{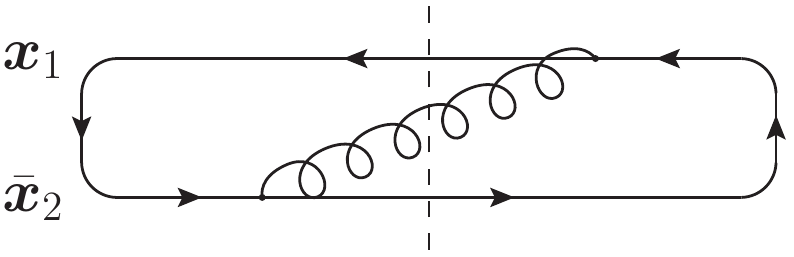}\\{\footnotesize (a)}
\end{center}
\end{minipage}
\begin{minipage}[b]{0.245\textwidth}
\begin{center}
\includegraphics[scale=0.5]{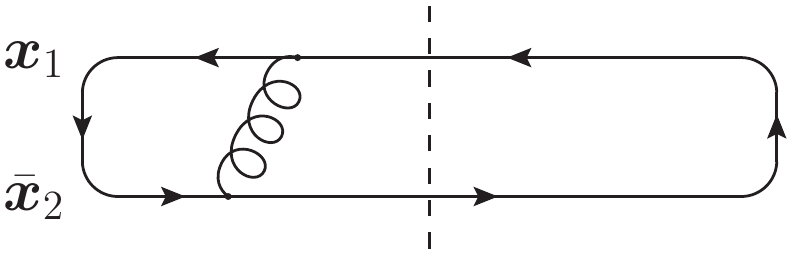}\\{\footnotesize (b)}
\end{center}
\end{minipage}
\begin{minipage}[b]{0.245\textwidth}
\begin{center}
\includegraphics[scale=0.5]{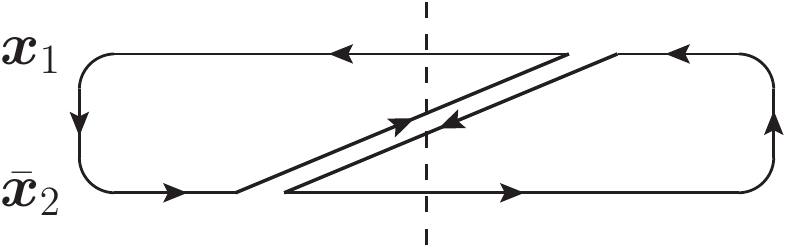}\\{\footnotesize (c)}
\end{center}
\end{minipage}
\begin{minipage}[b]{0.245\textwidth}
\begin{center}
\includegraphics[scale=0.5]{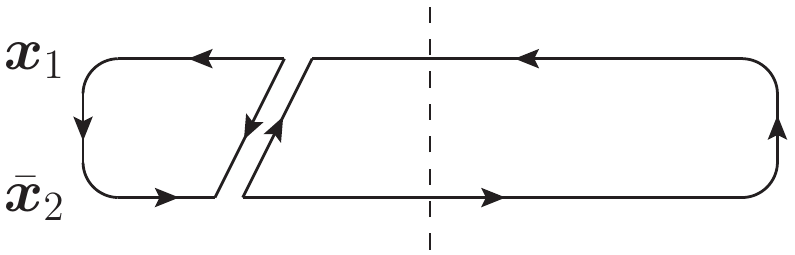}\\{\footnotesize (d)}
\end{center}
\end{minipage}
\caption{(a) A real emission of a gluon from a color dipole. (b) A virtual 
emission. (c) and (d) The corresponding diagrams in the large-$N_c$ limit. In all 
diagrams the dashed line stands for the interaction with the target.}
\label{fig:dipole}
\end{figure}

We shall mostly focus on the color dipole made by a quark-antiquark pair in an 
overall color singlet state, with the $S$-matrix
 \beq\label{Sdipole}
 \hat{S}_{\bmx_1\bmx_2} \equiv \hat{S}_{\bmx_1\bmx_2} ^{(2)}=
 \frac{1}{N_c}\,\rmtr({V}^{\dagger}_{\bmx_1} {V}_{\bmx_2}^{\phantom{\dagger}}),
 \eeq
and the color quadrupole, a system of two quarks and two antiquarks also in a color 
singlet, for which
 \beq\label{Squadrupole}
 \hat{Q}_{\bmx_1\bmx_2\bmx_3\bmx_4} \equiv
 \hat{S}_{\bmx_1\bmx_2\bmx_3\bmx_4}^{(4)}=
 \frac{1}{N_c}\,
 \rmtr({V}^{\dagger}_{\bmx_1} 
 {V}_{\bmx_2}^{\phantom{\dagger}}
 {V}^{\dagger}_{\bmx_3}
 {V}_{\bmx_4}^{\phantom{\dagger}}).
 \eeq
Here ${V}^{\dagger}$ and $V$ are Wilson lines like in \eqn{Vadj}, but in the 
fundamental representation. In general one can consider projectiles made with $n$ 
quarks and $n$ antiquarks, for which
\beq\label{S2n}
 \hat{S}_{\bmx_1\bmx_2 ...\bmx_{2n-1}\bmx_{2n}}^{(2n)} =
 \frac{1}{N_c}\,
 \rmtr({V}^{\dagger}_{\bmx_1} {V}_{\bmx_2}^{\phantom{\dagger}}\dots
 {V}^{\dagger}_{\bmx_{2n-1}}{V}_{\bmx_{2n}}^{\phantom{\dagger}}),
 \eeq
and eventually high energy evolution mixes such single-trace operators with 
multi-trace ones of the form
 \beq\label{multitrace}
 \hat{\mcal{O}} = \,\frac{1}{N_c}\,
 \rmtr({V}^{\dagger}_{\bmx_1} {V}_{\bmx_2}^{\phantom{\dagger}}\dots)
 \frac{1}{N_c}\,\rmtr({V}^{\dagger}_{\bmy_1} {V}_{\bmy_2}^{\phantom{\dagger}}\dots)
 \frac{1}{N_c}\,\rmtr({V}^{\dagger}_{\bmz_1} {V}_{\bmz_2}^{\phantom{\dagger}}
 \dots).
 \eeq
The quadrupole defined above in \eqn{Squadrupole} is the first non-trivial operator 
which can probe multi-parton correlation in the target wavefunction and will serve 
as the prototype in our study. In order to construct evolution equations according 
to Eqs.~\eqref{H} and \eqref{general}, it is necessary to specify how the 
functional derivatives w.r.t.~$\alpha_a$ act on observables. They act on endpoints 
of the Wilson lines according to
 \beq\label{donV}
 \frac{\delta}{\delta \alpha^a_{\bmu}}\,
 V_{\bmx}^{\dagger} =
 \rmi g \delta_{\bmx\bmu}\, t^a V_{\bmx}^{\dagger},
 \qquad
 \frac{\delta}{\delta \alpha^a_{\bmu}}\,
 V_{\bm{x}} =
 - \rmi g  \delta_{\bmx\bmu} V_{\bmx}\, t^a,
 \eeq
with $t^a$ in the fundamental representation and where we introduced the shorthand 
notation $\delta_{\bmx\bmu}=\delta^{(2)}(\bmx-\bmu)$. Because of their action on 
$V^{\dagger}$ (by convention), they are called ``left'' derivatives and could be 
also denoted as $\delta/\delta \alpha_{\rm L}^a$. Using these rules within 
Eqs.~\eqref{H} and \eqref{general}, it is straightforward to derive the evolution 
equations satisfied by the $S$-matrices for the dipole and the quadrupole. The 
resulting equation for the dipole is
 \beq\label{BK}
 \frac{\del \lan \hat{S}_{\bmx_1\bmx_2} \ran_Y}{\del Y}=
 \frac{\abar}{2\pi}\, \int_{\bmz}
 \mcal{M}_{\bmx_1\bmx_2\bmz}
 \lan \hat{S}_{\bmx_1\bmz} \hat{S}_{\bmz\bmx_2}
 -\hat{S}_{\bmx_1\bmx_2} \ran_Y,
 \eeq
where we have defined $\abar = \alpha_s N_c/\pi$. Even though derived by evolving 
the target, this equation has an easy interpretation in terms of projectile 
evolution, as shown in Fig.~\ref{fig:dipole}. The quadratic term in $\hat{S}$ has 
been generated by the real part of the Hamiltonian $H_{\rm real}$, that is, by the 
last two terms in the parenthesis in \eqn{H}. It describes the splitting of the 
original dipole $(\bmx_1,\bmx_2)$ into two new dipoles $(\bmx_1,\bmz)$ and 
$(\bmz,\bmx_2)$, which subsequently scatter off the target. More precisely, the 
evolution step consists of the emission of a soft gluon, hence the original dipole 
gets replaced by a quark-antiquark-gluon system, but in the large-$N_c$ limit this 
emission is equivalent to the aforementioned dipole splitting. The negative, linear 
in $\hat{S}$, term has been produced by the virtual part of the Hamiltonian 
$H_{\rm virt}$, that is, by the first two terms in the parenthesis in \eqn{H} and 
corresponds to the reduction in the probability for the dipole to survive in its 
original state. Notice that color transparency requires $\hat{S}_{\bmx\bmx}=1$ and 
thus the potential short-distance singularities, arising from the dipole kernel 
$\mcal{M}$, at $\bmx_1 = \bmz$ and $\bmx_2 = \bmz$ cancel between the real and virtual terms.

A word of caution should follow here. \eqn{BK} is valid for any value of $N_c$,
but still, both terms on the right hand side are of the same order in $N_c$. 
In fact, terms suppressed by $1/N_c^2$ have been generated in intermediate steps of the calculation but they have canceled in the final result. More precisely, when acting with 
$H_{\rm real}$ on $\hat{S}_{\bmx_1\bmx_2}$ we get a contribution
 \beq
 \label{real}
 -\frac{1}{N_c^2}\,  
 \atpi
 \int_{\bm{z}}
 \mcal{M}_{\bmx_1\bmx_2\bmz}
 \hat{S}_{\bmx_1\bmx_2},
 \eeq
which cancels with an opposite in sign contribution coming from the action of $H_{\rm virt}$.

\begin{figure}
\begin{minipage}[b]{0.40\textwidth}
\begin{center}
\includegraphics[scale=0.55]{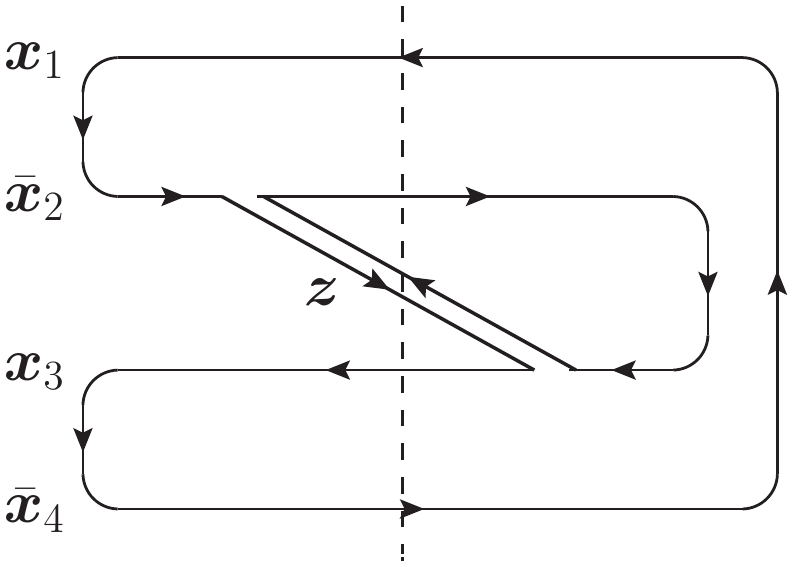}\\{\footnotesize (a)}
\end{center}
\end{minipage}
\begin{minipage}[b]{0.40\textwidth}
\begin{center}
\includegraphics[scale=0.55]{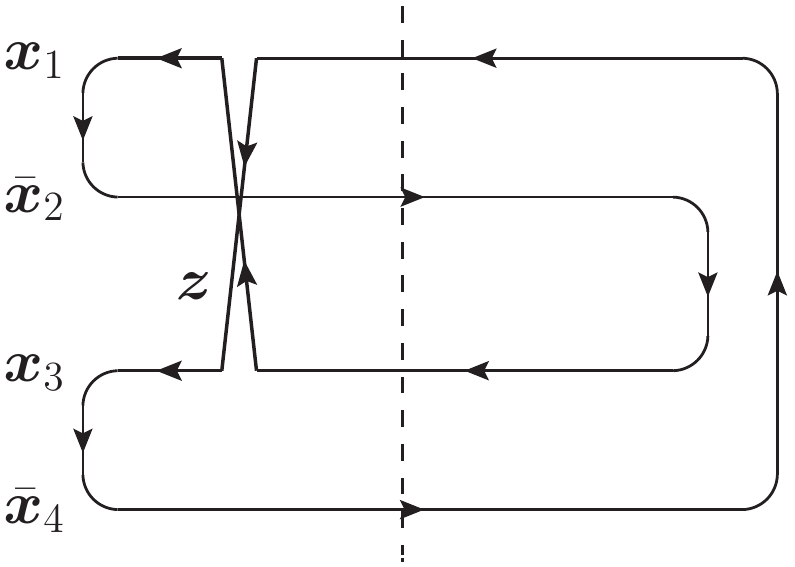}\\{\footnotesize (b)}
\end{center}
\end{minipage}
\caption{(a) A real emission of a gluon from a color quadrupole in the large-$N_c$ 
limit corresponding to the term proportional to 
$\mcal{M}_{\bmx_2\bmx_3\bmz} \hat{S}_{\bmz\bmx_2}\hat{Q}_{\bmx_1\bmz\bmx_3\bmx_4}$. 
(b) A virtual emission corresponding to the term proportional 
to $-\mcal{M}_{\bmx_1\bmx_3\bmz}\hat{S}_{\bmx_3\bmx_2}\hat{S}_{\bmx_1\bmx_4}$. In 
both diagrams the dashed line denotes the interaction with the target.}
\label{fig:quadrupole}
\end{figure}

Similarly one can derive the evolution equation for the quadrupole, which reads
 \begin{align}\label{Qevol}
 \hspace{-0.2cm}
 \frac{\del \lan\hat{Q}_{\bmx_1\bmx_2\bmx_3\bmx_4} \ran_Y}{\del Y} =
 \frac{\abar}{4\pi}  \int_{\bmz} &
 \Big[(\mcal{M}_{\bmx_1\bmx_2\bmz} +
 \mcal{M}_{\bmx_1\bmx_4\bmz} -
 \mcal{M}_{\bmx_2\bmx_4\bmz})
 \lan
 \hat{S}_{\bmx_1\bmz}\hat{Q}_{\bmz\bmx_2\bmx_3\bmx_4}
 \ran_Y
 \nn
 &+(\mcal{M}_{\bmx_1\bmx_2\bmz} +
 \mcal{M}_{\bmx_2\bmx_3\bmz} -
 \mcal{M}_{\bmx_1\bmx_3\bmz})
 \lan
 \hat{S}_{\bmz\bmx_2}\hat{Q}_{\bmx_1\bmz\bmx_3\bmx_4}
 \ran_Y
 \nn
 &+(\mcal{M}_{\bmx_2\bmx_3\bmz} +
 \mcal{M}_{\bmx_3\bmx_4\bmz} -
 \mcal{M}_{\bmx_2\bmx_4\bmz})
 \lan
 \hat{S}_{\bmx_3\bmz}\hat{Q}_{\bmx_1\bmx_2\bmz\bmx_4}
 \ran_Y
 \nn
 &+(\mcal{M}_{\bmx_1\bmx_4\bmz} +
 \mcal{M}_{\bmx_3\bmx_4\bmz} -
 \mcal{M}_{\bmx_1\bmx_3\bmz})
 \lan
 \hat{S}_{\bmz\bmx_4}\hat{Q}_{\bmx_1\bmx_2\bmx_3\bmz}
 \ran_Y
 \nn
 &-(\mcal{M}_{\bmx_1\bmx_2\bmz} + \mcal{M}_{\bmx_3\bmx_4\bmz}
 +\mcal{M}_{\bmx_1\bmx_4\bmz} + \mcal{M}_{\bmx_2\bmx_3\bmz})
 \lan
 \hat{Q}_{\bmx_1\bmx_2\bmx_3\bmx_4}
 \ran_Y
 \nn
 &-(\mcal{M}_{\bmx_1\bmx_2\bmz} + \mcal{M}_{\bmx_3\bmx_4\bmz}
 -\mcal{M}_{\bmx_1\bmx_3\bmz} - \mcal{M}_{\bmx_2\bmx_4\bmz})
 \lan
 \hat{S}_{\bmx_1\bmx_2}\hat{S}_{\bmx_3\bmx_4}
 \ran_Y
 \nn
 &-(\mcal{M}_{\bmx_1\bmx_4\bmz} + \mcal{M}_{\bmx_2\bmx_3\bmz}
 -\mcal{M}_{\bmx_1\bmx_3\bmz} - \mcal{M}_{\bmx_2\bmx_4\bmz})
 \lan
 \hat{S}_{\bmx_3\bmx_2}\hat{S}_{\bmx_1\bmx_4}
 \ran_Y\Big].
 \end{align}
Even though this looks considerably more involved than the dipole equation, a
similar discussion applies and two representative diagrams are shown in 
Fig.~\ref{fig:quadrupole}. The terms involving $\lan \hat{S} \hat{Q}\ran_Y$ in the 
right hand side are real terms describing the splitting of the original quadrupole 
into a new quadrupole plus a dipole, and have generated by the action 
$H_{\rm real}$. The virtual terms involving $\lan \hat{Q}\ran_Y$ and 
$\lan \hat{S} \hat{S}\ran_Y$ are necessary for probability conservation, and have 
been generated by $H_{\rm virt}$. Once again, all terms subleading at large $N_c$, 
separately generated by the two parts of the Hamiltonian, have canceled in the 
final equation, and all the short-distance singularities of the dipole kernels at 
$\bmx_i=\bmz$, with $i=1,2,3,4$, cancel among the various terms.

All the above features generalize to the evolution equations obeyed by the 
single-trace observables given in \eqn{S2n}. As already evident in Eqs.~\eqref{BK} 
and \eqref{Qevol}, these equations are generally not closed and they couple 
single-trace observables with multi-trace ones. For instance, the quadrupole 
equation involves the 4-point function $\lan \hat{S} \hat{S}\ran_Y$ and the 6-point 
function $\lan \hat{S} \hat{Q}\ran_Y$, which in turn will couple to even 
higher-point correlators. The equations obeyed by the multi-trace observables 
exhibit the new feature that they involve explicit $1/N_c^2$ corrections. These 
arise when two functional derivatives in \eqn{H} act on Wilson lines which belong 
to different traces (see e.g.~Appendix F in \cite{Triantafyllopoulos:2005cn} for an 
example). At large $N_c$, one can neglect these terms and check that the hierarchy 
admits the factorized solution 
 \beq\label{fact}
 \lan\hat{\mcal{O}}\ran_Y \simeq
 \Big\lan\frac{1}{N_c}\,\rmtr({V}^{\dagger}_{\bmx_1}
 {V}_{\bmx_2}^{\phantom{\dagger}}\dots)\Big\ran_Y
 \Big\lan\frac{1}{N_c}\,\rmtr({V}^{\dagger}_{\bmy_1}
  {V}_{\bmy_2}^{\phantom{\dagger}}\dots)\Big\ran_Y
 \Big\lan\frac{1}{N_c}\,\rmtr({V}^{\dagger}_{\bmz_1}
  {V}_{\bmz_2}^{\phantom{\dagger}}\dots)\Big\ran_Y\dots\,,
 \eeq
so long as this factorization is already present in the initial condition. 
Therefore the hierarchy simplifies in a drastic way as it decomposes 
into a set of equations which can be solved, in principle, one after the other. 
More precisely, \eqn{BK} becomes a closed equation for $\lan\hat{S}\ran_Y$, 
the well-known BK equation \cite{Balitsky:1995ub,Kovchegov:1999yj}, \eqn{Qevol} 
becomes an inhomogeneous equation for $\lan \hat{Q} \ran_Y$ whose coefficients 
depend on $\lan\hat{S}\ran_Y$ \cite{JalilianMarian:2004da}, and so on. 
Good analytical understanding and reliable numerical solutions to the BK equation 
exist, and we will discuss these matters in the next section. However, already 
starting from the quadrupole, it seems difficult to numerically deal with the large 
number of variables in the higher-point equations and the non-locality in the 
transverse space at the same time. In that case one can rely on an alternative 
formulation of the JIMWLK evolution as a Langevin equation \cite{Blaizot:2002xy} 
which can be solved on a lattice 
\cite{Rummukainen:2003ns,Lappi:2011ju,Dumitru:2011vk}, or 
develop well-motivated approximate schemes and we shall turn our attention to the 
latter in Sect.~\ref{sec:gauss}.

\section{\label{sec:bk}Solution to the BK equation and the Levin-Tuchin formula}

Let us briefly review the analytic solution to the BK equation. We shall assume a 
homogeneous target so that the amplitude 
$\lan \hat{T}_{\bmx_1\bmx_2} \ran_Y = 1 - \lan\hat{S}_{\bmx_1\bmx_2}\ran_Y$ 
depends only on the magnitude $r \equiv r_{12} = |\bmx_1-\bmx_2|$. Still, the 
solution is not analytically known, but one can construct a piecewise one when 
$\abar Y \gtrsim 1$, by considering two regimes. The regime $r \lesssim 1/Q_s$ 
where the target is dilute and the scattering weak and the regime $r \gtrsim 1/Q_s$ 
where the target is dense and the scattering strong. The borderline in between 
corresponds to the saturation momentum $Q_s$ and can be determined from the 
solution to the BFKL equation, that is, the linearized in $\hat{T}$ version of the 
BK equation, supplemented by appropriate boundary conditions. For fixed coupling 
one finds that the energy dependence of the saturation momentum is determined 
by \cite{Mueller:2002zm,Munier:2003sj}
 \beq\label{lambda}
 \frac{1}{\abar}\,
 \frac{\dif \ln Q_s^2}{\dif Y} = 
 \frac{\chi(\gamma_s)}{\gamma_s}
 -\frac{3}{2 \gamma_s}\,\frac{1}{\abar Y}, 
 \eeq 
where the ``anomalous dimension'' $\gamma_s$ related to saturation is determined by \cite{Gribov:1984tu}
 \beq
 \chi'(\gamma_s) = \chi(\gamma_s)/\gamma_s
 \,\Rightarrow\, \gamma_s \approx 0.628.
 \eeq
In the above, $\chi(\gamma)$ is the eigenvalue function of the BFKL 
equation \cite{Kuraev:1977fs,Balitsky:1978ic} given by
 \beq
 \chi(\gamma) = 2 \psi(1) - \psi(\gamma) - \psi(1-\gamma),
 \eeq
with $\psi$ the logarithmic derivative of the $\Gamma$-function. The amplitude 
on this side of the saturation line, that is for $r \ll 1/Q_s$, reads \cite{Mueller:2002zm,Munier:2003sj}
 \beq \label{Tweak}
 \lan \hat{T} \ran_Y = (r^2 Q_s^2)^{\gamma_s}
 \ln \left(\frac{1}{r^2 Q_s^2} + c \right)
 \exp \left[-\frac{\ln^2(r^2 Q_s^2)}{D_s \abar Y} \right],
 \eeq
an expression which is valid in the region 
$Q_s^2 \ll 1/r^2 \ll Q_s^2 \exp(D_s \abar Y)$, where $c$ is a constant of 
order $\order{1}$ and $D_s = 2 \chi''(\gamma_s) \approx 97$ is the diffusion coefficient. 
When $|\ln(r^2 Q_s^2)| \ll \sqrt{D_s \abar Y}$ the last factor in \eqn{Tweak}, 
which describes diffusion, can be set equal to unity and the amplitude exhibits 
geometrical scaling \cite{Stasto:2000er,Iancu:2002tr,Mueller:2002zm,Munier:2003vc,Munier:2003sj}; 
it depends only on the combined variable $r^2 Q_s^2$.

Let us now look at what happens when $r \gg 1/Q_s$. In this regime the 
$S$-matrix approaches its black-disk limit, i.e.~$\lan \hat{S} \ran_Y \to 0$, and 
thus we can neglect the term quadratic in $\lan \hat{S} \ran_Y$ in the BK equation. 
To do this properly, we need to restrict the region of integration
in the transverse coordinates to $1/Q_s^2 \ll |\bmx_i-\bmz|^2\ll r^2$, 
with $i=1,2$. The lower limit emerges as the boundary determining the transition 
from a region where the scattering is weak to a region
where it becomes strong. The upper limit determines the dominant logarithmic 
contribution to the r.h.s.~of the BK equation. We easily find that
 \beq\label{BKsat}
 \frac{\del \lan \hat{S}\ran_Y}{\del Y} = 
 -\abar \int_{1/Q_s^2}^{r^2}\frac{\dif z^2}{z^2}\, \lan \hat{S}\ran_Y = 
 -\abar \ln(r^2 Q_s^2)\, \lan \hat{S}\ran_Y,
 \eeq
and using the leading behavior for the rapidity dependence of $Q_s$ given 
in \eqn{lambda} we convert the derivative w.r.t.~Y to one w.r.t.~$\ln(r^2 Q_s^2)$. 
Then it becomes trivial to solve \eqn{BKsat} and we arrive at
 \beq\label{Ssat}
 \lan \hat{S} \ran_Y \sim 
 \exp\left[-\frac{\gamma_s}{2 \chi(\gamma_s)}\,\ln^2(r^2 Q_s^2)\right], 
 \eeq
where it should be reminded that only this leading term in the exponent is under 
good control. \eqn{Ssat} is equivalent in its functional form to the 
Levin-Tuchin formula 
\cite{Levin:1999mw} (see also \cite{Mueller:1996te,Iancu:2001md,Mueller:2002pi,Iancu:2003zr}), apart from the coefficient in the exponent which is 
different. Let us also add here, that this coefficient is further modified when 
$N_c$ is finite, namely the 
exponent should be multiplied by a factor $2 C_F/N_c = (N_c^2 - 1)/ N_c^2$ 
\cite{Mueller:2002pi,Iancu:2011nj}. Again, geometric scaling is manifest in 
\eqn{Ssat} and in fact this property is valid everywhere in the region 
$\Lambda^2_{\rm QCD} \ll 1/r^2 \ll Q_s^2 \exp[\sqrt{D_s \abar Y}]$. Even though we 
do not have an analytical solution to cover the whole region, one can construct 
appropriate interpolations. For instance, and neglecting also the second factor in 
\eqn{Tweak} which has only a weak logarithmic dependence, such a convenient 
expression is
 \beq
 \lan \hat{S} \ran_Y = 
 \exp\left\{-\frac{2}{\gamma_s \chi(\gamma_s)}\,
 \ln^2 \big[1 + b (r Q_s)^{\gamma_s}\big]\right\},
 \eeq
with $b$ a constant of order $\order{1}$.

Going back to \eqn{Ssat} and using \eqn{lambda}, let us note that, for fixed $r$, 
the dominant term in the exponent is proportional to the square of $Y$. This has a 
direct physical interpretation which should have been obvious from the derivation. 
One factor of $Y$ already appears in zero-dimensional particle models where all 
transverse coordinates are suppressed and represents the fact that $\dif P/\dif Y$ 
is proportional to $-P$, where $P$ is the probability that a particle does not 
split. An extra factor of $Y$ arises in QCD because the available phase space for 
an emission of  a gluon increases with $\ln(Q_s^2) \sim Y$.

\begin{figure}[t]
\begin{center}
\includegraphics[width=1\textwidth,angle=0]{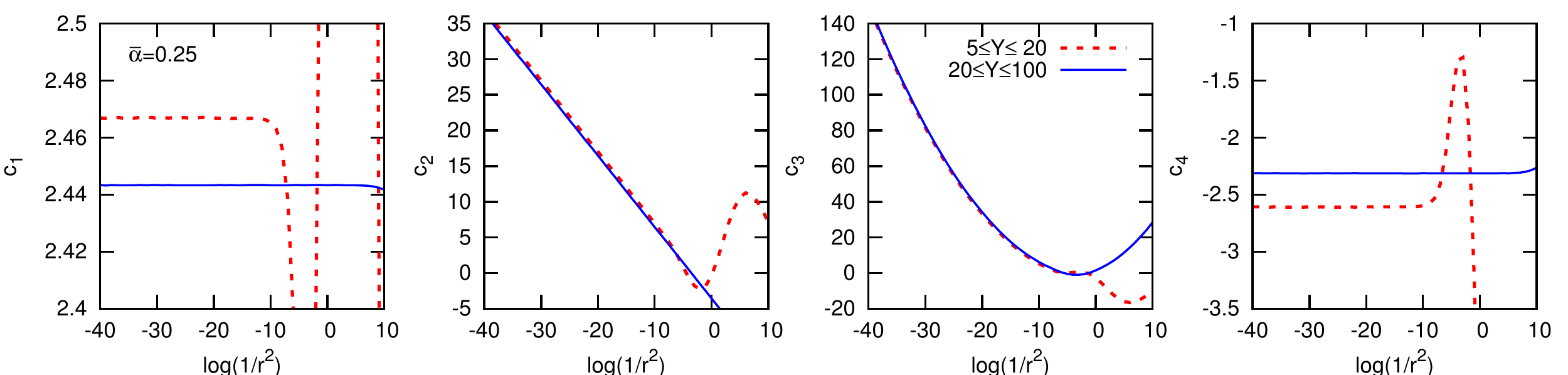}
\end{center}
\caption{Numerical extraction of the coefficients in the asymptotic
  expansion \eqref{eq:ltfit} at saturation and for fixed coupling. The
coefficients are plotted as a function of the dipole size $r$.}
\label{fig:LTfc}
\end{figure}

Here we shall put forward the task to numerically verify \eqn{Ssat} for two 
reasons. The first is simply that such a study has not been done so far 
(see \cite{Mueller:1996te} for a related work). The second is that it is necessary 
for the comparisons to be performed in Sect.~\ref{sec:mfa}. To this end, by also 
using \eqn{lambda}, we can write the large-$Y$ and large-$\ln(1/r^2)$ expansion 
 \beq\label{lnS}
 -\ln \lan \hat{S} \ran_Y 
 = \frac{\chi(\gamma_s)}{2\gamma_s}\, (\abar Y)^2
   - \ln\frac{1}{r^2}\, (\abar Y)
 + \frac{\gamma_s}{2 \chi(\gamma_s)} \,\ln^2\frac{1}{r^2}
 - \frac{3}{2\gamma_s}\, (\abar Y) \ln(\abar Y).
 \eeq
We compare this expression with fits of the rapidity dependence of
numerical solutions of the BK equation\footnote{Details about the
  numerical implementation of the BK equation and the other evolution
  equations considered later in this article are given in appendix.}
with a function of the form
 \beq\label{eq:ltfit}
 -\ln \lan \hat{S} \ran_Y 
 = c_1\, (\abar Y)^2
 + c_2\, (\abar Y)
 + c_3
 + c_4\, (\abar Y) \ln(\abar Y).
 \eeq
The excellent numerical verification of all terms in \eqref{lnS} is demonstrated in Fig.~\ref{fig:LTfc} and in Table \ref{tab:cfc}. This makes clear that \eqn{Ssat}, including the exact coefficient in the exponent, gives the correct approach to the unitarity limit within the context of the BK equation. Notice that the asymptotics sets quite fast and the Levin-Tuchin law is valid also for ``reasonable'' values of $Y$, albeit this happens for rather large dipoles. Furthermore, it is remarkable that, without focusing in the transition region around $Q_s$, we get a numerical confirmation for the first two terms of its energy dependence as given in \eqn{lambda}.

\begin{table}[t!]
\begin{center}
\begin{tabular}{ | l || c | c | c | c |}
\hline
Coefficient 
& $c_1$      
& $c_2/\ln(1/r^2)$  
& $c_3/\ln^2(1/r^2)$
& $c_4$ 
\\ \hline \hline
Analytical 
& $\chi(\gamma_s)/2\gamma_s = 2.442$ 
& $-1$ 
& $\gamma_s/2 \chi(\gamma_s) = 0.1024$
& $-3/2\gamma_s = -2.390$
\\ \hline
Numerical ($5<Y<20$)  
& 2.466
& -1.002 
& 0.1029
& -2.601
\\ \hline
Numerical ($20<Y<200$) 
& 2.443 
& -1.000 
& 0.1045 
& -2.313
\\ \hline
\end{tabular}
\end{center}
\caption{\label{tab:cfc} Analytical and numerical values of the coefficients in the 
asymptotic expansion of $-\ln \lan \hat{S}(r) \ran_Y$ at saturation as in 
Fig.~\ref{fig:LTfc}. The agreement of the numerics in Fig.~\ref{fig:LTfc} 
with \eqn{lnS} is excellent.}
\end{table} 

\begin{figure}
\begin{center}
\includegraphics[width=0.85\textwidth,angle=0]{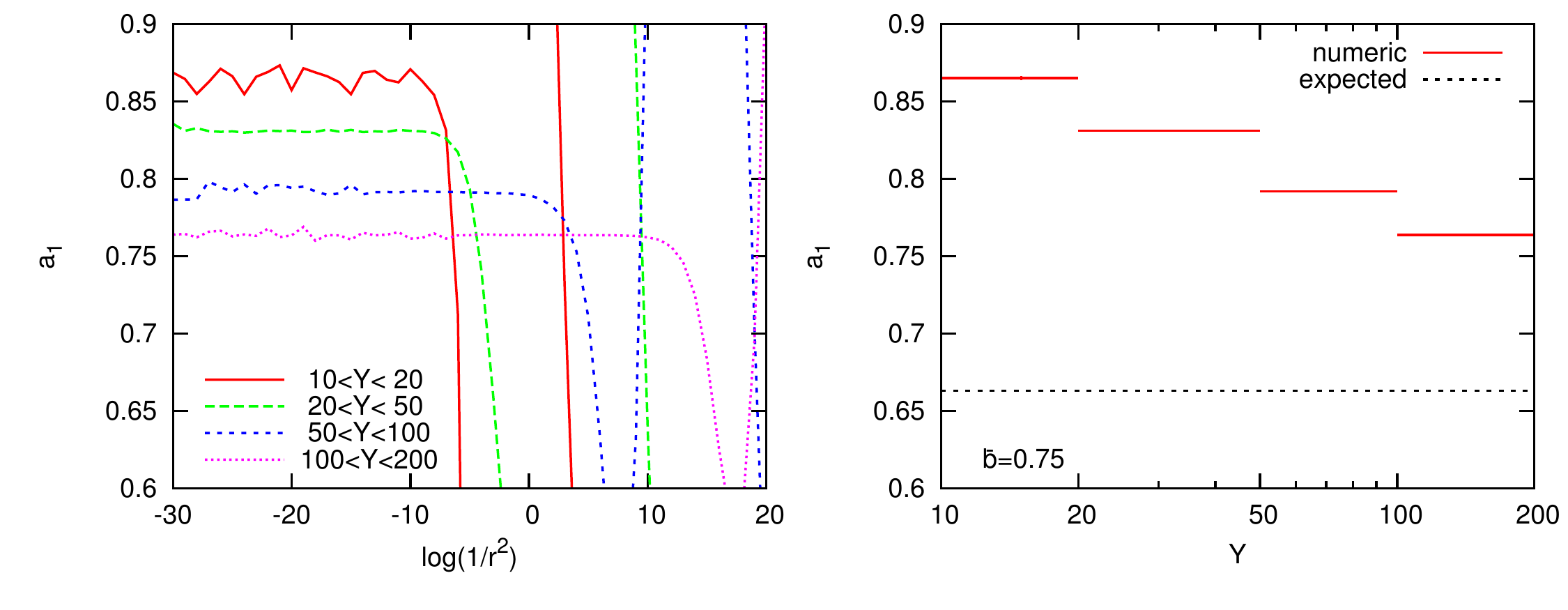}
\end{center}
\caption{Numerical extraction of the coefficient $a_1$ in the asymptotic expansion  
$-\ln \lan \hat{S}(r) \ran_Y = a_1 Y \ln Y + a_2  Y + a_3 \ln Y + a_4$ at 
saturation and for running coupling.}
\label{fig:LTrc}
\end{figure}

It is not hard to see how the above description is modified when we consider the
running coupling BK equation \cite{Kovchegov:2006vj,Balitsky:2006wa,Balitsky:2008zza}. In general, the coupling runs according to
 \beq
 \abar(r) = \frac{1}{\bar{b} \ln(1/r^2 \Lambda^2_{\rm QCD})}
 \quad \mathrm{with} \quad
 \bar{b} =\frac{11 N_c - 2 N_f}{12 N_c},
 \eeq
and where one would need an appropriate prescription to freeze it when $r$ starts 
to approach $1/\Lambda_{\rm QCD}$. Now the energy dependence of the saturation 
momentum reads \cite{Mueller:2002zm,Munier:2003sj,Triantafyllopoulos:2002nz,Beuf:2010aw}
 \beq
 \frac{\dif \ln Q_s^2}{\dif Y} = 
 \sqrt{\frac{\chi(\gamma_s)}{2 \bar{b} \gamma_s}\,\frac{1}{Y}}
 - \frac{|\xi_1|}{8} 
 \left[\frac{\chi''(\gamma_s)}{\chi(\gamma_s)}\right]^{1/3}
 \left[\frac{\chi(\gamma_s)}{2\bar{b} \gamma_s}\right]^{1/6}
 \frac{1}{Y^{5/6}},
 \eeq
with $\xi_1 = -2.33$ the leftmost zero of the Airy function. In order to find the behavior of the $S$-matrix at saturation, we make in \eqn{BKsat} the replacement
 \beq
 \abar \to \frac{1}{\bar{b} \ln(1/z^2 \Lambda^2_{\rm QCD})},
 \eeq
and where clearly this factor has to be moved inside the integrand. Notice that 
such a replacement can be obtained from a ``smallest dipole'' prescription, which 
is a natural one\footnote{In the subsequent sections we shall extend our study to include two more prescriptions: the Balitsky prescription \cite{Balitsky:2006wa} which has dominated the phenomenology in the recent years, and a particular daughter dipole prescription which is used when one solves the Langevin form of the JIMWLK equation \cite{Lappi:2011ju,Dumitru:2011vk}.}; given the dipole splitting 
$(\bmx_1,\bmx_2)$ $\to$ $(\bmx_1,\bmz)$ and $(\bmz,\bmx_2)$ we  let the coupling 
run according to $\abar(r_{\rm min})$, with $r_{\rm min}$ the size of the smallest 
of the three dipoles involved in the process. We find \cite{Mueller:2002pi}
 \beq
 \frac{\del \lan \hat{S}\ran_Y}{\del Y} = 
 -\frac{1}{\bar{b}}\,
 \ln\left[\frac{\ln(Q_s^2/\Lambda^2_{\rm QCD})}{\ln(1/r^2 \Lambda^2_{\rm QCD})}\right]\,
 \lan \hat{S}\ran_Y,
 \eeq
and after the integration by keeping only the most dominant term in $Y$ we obtain
 \beq\label{SsatYrun}
 \lan \hat{S} \ran_Y \sim 
 \exp\left(-\frac{1}{2 \bar{b}}\,Y \ln Y\right).
 \eeq
In the numerical solution shown in Fig.~\ref{fig:LTrc} we see that one indeed 
recovers this dependence in $Y$ without the need to evolve too much. However the 
value of the coefficient differs around 30\% from its asymptotic value 
$1/2\bar{b}$. The approach to the latter is extremely slow and takes place 
only at tremendously high values of $Y$. 

\section{\label{sec:gauss}The Gaussian approximation}

As said at the end of Sect.~\ref{sec:jimwlk}, one method to calculate multi-gluon 
correlators is to solve the Langevin form of the JIMWLK equation on a lattice 
\cite{Blaizot:2002xy}. The implementation of this approach \cite{Dumitru:2011vk} 
lead to the remarkable finding that the numerical data for the quadrupole are 
well-described by a Gaussian approximation, more precisely by the
extension of the MV model to arbitrary rapidity $Y$. We shall not review here the 
MV model, but is suffices for our purposes to recall that it is the typical initial 
condition at some $Y_0$. It is a Gaussian functional in the color sources $\rho_a$ 
and therefore also in the field $\alpha_a$ in the covariant gauge, but with a 
kernel which is independent of $Y$, since it refers to a fixed, initial, value of 
rapidity. Furthermore, it is clear that any Gaussian probability distribution 
involves only a single kernel, thus all high-point correlators can be expressed in 
terms of the 2-point one, e.g.~the dipole $S$-matrix $\lan \hat{S} \ran_Y$, a 
property which is true in the MV-model. Nevertheless, there was no a priori reason 
for this to happen for the JIMWLK Hamiltonian which is highly non-linear due to the 
Wilson lines in \eqn{H} and which arise, as said, from the scattering of the 
emitted gluon with the exisiting background target field. Perhaps a hint was given 
some time ago in \cite{Iancu:2002aq}, where a ``random phase approximation'' to 
JIMWLK lead to a Gaussian Hamlitonian, even though no explicit reference to 
high-point correlators was made there. A Gaussian ansatz has also been used in 
\cite{Kovchegov:2008mk} in order to estimate $1/N_c^2$ corrections to the BK 
equation.

Still, despite the aforementioned nonlinearities, one can prove that a Gaussian 
approximation is a quasi-exact solution to the JIMWLK equation 
\cite{Iancu:2011ns,Iancu:2011nj}. The non-linearity is indeed present in the 
dipole equation given in \eqn{BK}, but not in the quadrupole one, \eqn{Qevol}, 
which is linear in $\hat{Q}$ (at the operator level and/or at large-$N_c$). This is 
an indication that a Gaussian approximation may be a possible solution with all the 
non-linearities absorbed in its kernel or, equivalently, in the 2-point function. 
Let us shortly review and explain how this happens 
\cite{Iancu:2011ns,Iancu:2011nj}.

At saturation, where by definition the target is dense, real emissions are 
suppressed and the virtual part $H_{\rm virt}$ of the Hamiltonian dominates. This 
part has obviously a Gaussian form, including the second term; those adjoint 
Wilson lines simply transform the ``left'' functional derivatives to ``right'' ones 
which act on the lower and upper end-points of the Wilson lines $V^{\dagger}$ and 
$V$ respectively, namely
 \beq\label{dRonV}
 (\wt{V}_{\bmu}^{\dagger})^{ac}
 \frac{\delta}{\delta \alpha^a_{\bmu}}\,
 V_{\bmx}^{\dagger} =
 (\wt{V}_{\bmu}^{\dagger})^{cb}
 \frac{\delta}{\delta \alpha^b_{\bmu}}\,
 V_{\bmx}^{\dagger}
 \equiv
 \frac{\delta}{\delta \alpha_{{\rm R} {\bmu}}^a}\,
 V_{\bmx}^{\dagger}=
 \rmi g \delta_{\bmx\bmu} V_{\bmx}^{\dagger} t^c,
 \eeq
and similarly for the action on $V_{\bmx}^{\phantom{\dagger}}$.

Now we proceed as in the case of the BK equation. Since the integrand of 
$H_{\rm virt}$ is $z$-independent, we can integrate the dipole kernel over $\bmz$ 
in the region $1/Q_s^2 \ll |\bmu-\bmz|^2,|\bmv-\bmz|^2 \ll |\bmu-\bmv|^2$. Again 
the lower limit is imposed by our approximation, while the upper one is chosen to 
give the dominant logarithmic contribution which is 
$2 \ln\big[(\bmu-\bmv)^2 Q_s^2\big]$. So far our approach would lead to a 
Hamiltonian valid only at saturation, but recalling \eqn{BKsat} we see that this 
logarithm can be expressed in terms of the logarithmic derivative of the dipole 
w.r.t.~$Y$ and with such a replacement we finally 
arrive at
 \beq\label{HG}
 H_{\rm G} = \frac{1}{4 g^2 C_F}
 \int_{\bmu\bmv}
 \frac{\dif \ln \blan \hat{S}_{\bmu\bmv}\bran_Y}{\dif Y}\,
 \left(\frac{\delta}{\delta \alpha_{{\rm L}\bmu}^a} 
 \frac{\delta}{\delta \alpha_{{\rm L}\bmv}^a}
 +
 \frac{\delta}{\delta \alpha_{{\rm R}\bmu}^a} 
 \frac{\delta}{\delta \alpha_{{\rm R}\bmv}^a}
 \right).
 \eeq
This is a Gaussian Hamiltonian which is correct, for finite-$N_c$, at
saturation by construction and in the dilute limit as can be inspected. The kernel, 
which has absorbed the non-linearities, is $Y$-dependent (in contrast to the MV 
model one) and is most easily determined from the BK equation. Indeed one can 
verify that in the Gaussian approximation, and in an arbitrary representation $R$, 
we have
 \beq\label{Srel} 
 \ln \blan \hat{S}^{R}_{\bmu\bmv}\bran_Y 
 = \frac{2 C_R}{N_c}\, 
 \ln \blan \hat{S}^{\rm BK}_{\bmu\bmv}\bran_Y,
 \eeq 
and insert it for $R=F$ in \eqn{HG}. Therefore, by solving the BK equation, a 
closed equation at large-$N_c$, we can first find the dipole $S$-matrix at 
finite-$N_c$ using \eqn{Srel}. Then, using $H_{\rm G}$ one constructs evolution 
equations for high-point correlators which have the advantage to be local in the 
transverse plane, which simply means they are ordinary differential equations in 
$Y$ with $Y$-dependent coefficients. For example, the quadrupole at large-$N_c$, on which we will focus 
in Sect.~\ref{sec:quad}, reads \cite{Iancu:2011ns,Iancu:2011nj}
 \begin{align}
 \label{QconvS}
 &\hspace*{-1cm}
 \lan \hat{Q}_{\bmx_1\bmx_2\bmx_3\bmx_4} \ran_Y =
 \sqrt{\lan \hat{S}_{\bmx_1\bmx_2} \ran_Y
 \lan \hat{S}_{\bmx_3\bmx_2} \ran_Y
 \lan \hat{S}_{\bmx_3\bmx_4} \ran_Y
 \lan \hat{S}_{\bmx_1\bmx_4} \ran_Y
 }
 \left[
 \frac{\lan\hat{Q}_{\bmx_1\bmx_2\bmx_3\bmx_4} \ran_{Y_0}}
 {\sqrt{\lan \hat{S}_{\bmx_1\bmx_2} \ran_{Y_0}
 \lan \hat{S}_{\bmx_3\bmx_2} \ran_{Y_0}
 \lan \hat{S}_{\bmx_3\bmx_4} \ran_{Y_0}
 \lan \hat{S}_{\bmx_1\bmx_4} \ran_{Y_0}
 }} \right.
 \nn
 &\hspace*{-1cm} \left.+ \frac{1}{2}\int_{Y_0}^{Y} \dif y\,
 \frac{\lan \hat{S}_{\bmx_1\bmx_3} \ran_y
 \lan \hat{S}_{\bmx_2\bmx_4} \ran_y}
 {\sqrt{\lan \hat{S}_{\bmx_1\bmx_2} \ran_y
 \lan \hat{S}_{\bmx_3\bmx_2} \ran_y
 \lan \hat{S}_{\bmx_3\bmx_4} \ran_y
 \lan \hat{S}_{\bmx_1\bmx_4} \ran_y
 }}\,
 \frac{\del}{\del y}
 \frac{\lan \hat{S}_{\bmx_1\bmx_2} \ran_y
 \lan \hat{S}_{\bmx_3\bmx_4} \ran_y+
 \lan \hat{S}_{\bmx_1\bmx_4} \ran_y
 \lan \hat{S}_{\bmx_3\bmx_2} \ran_y}{\lan \hat{S}_{\bmx_1\bmx_3} \ran_y
 \lan \hat{S}_{\bmx_2\bmx_4} \ran_y}\right]\!.
 \end{align}
The  quadrupole above obeys the ``mirror symmetry'' 
$\blan\hat{Q}_{\bmx_1\bmx_2\bmx_3\bmx_4} \bran_Y = 
\blan\hat{Q}_{\bmx_1\bmx_4\bmx_3\bmx_2} \bran_Y$, a property which in fact holds at 
finite-$N_c$ and beyond the Gaussian approximation and is a consequence of 
symmetry under time-reversal where time is represented by $x^{-}$ \cite{Iancu:2011nj}. This symmetry is preserved by the JIMWLK equation due to the two types, left and right, of functional derivatives and suggests that the hadron expands symmetrically in the $x^{-}$ direction during the evolution. At the level  of the Gaussian approximation only, the quadrupole is also symmetric under the charge conjugation 
$\blan\hat{Q}_{\bmx_1\bmx_2\bmx_3\bmx_4} \bran_Y = 
\blan\hat{Q}_{\bmx_2\bmx_3\bmx_4\bmx_1} \bran_Y$.
It is also important to check the weak scattering limit of \eqn{QconvS}; when all $\lan \hat{T}_{\bmx_i\bmx_j} \ran_Y = 
1 - \lan \hat{S}_{\bmx_i\bmx_j} \ran_Y$ are small, then the expansion of \eqn{QconvS} 
to first order in $\lan \hat{T}\ran_Y$ reads
 \beq
 \label{Qweak}
 \lan \hat{Q}_{\bmx_1\bmx_2\bmx_3\bmx_4} \ran_Y = 1 
 - \lan \hat{T}_{\bmx_1\bmx_2} \ran_Y
 + \lan \hat{T}_{\bmx_1\bmx_3} \ran_Y
 - \lan \hat{T}_{\bmx_1\bmx_4} \ran_Y
 - \lan \hat{T}_{\bmx_2\bmx_3} \ran_Y
 + \lan \hat{T}_{\bmx_2\bmx_4} \ran_Y
 - \lan \hat{T}_{\bmx_3\bmx_4} \ran_Y,
 \eeq
which is assuredly the same relation one finds by expanding the Wilson lines of the 
dipole and the quadrupole to order $(g\alpha)^2$. Let us also point
out that a general initial condition where $\lan \hat{Q} \ran_{Y_0}$
is not determined by $\lan \hat{S} \ran_{Y_0}$ like, for example, the
one in \cite{Dumitru:2011ax}, can be accommodated for the quadrupole
in \eqn{QconvS}.

In some scenarios one can perform analytically the $y$-integration in an equation like \eqn{QconvS} and therefore arrive at a functional expression for a high-point correlator in terms of the dipole which is local in $Y$. Such a situation is realized for all the simple configurations studied in \cite{Dumitru:2011vk,Iancu:2011ns,Iancu:2011nj} and independently of whether the coupling is running or not. Furthermore, using a ``separability'' property of the 
Gaussian kernel \cite{Iancu:2011ns,Iancu:2011nj} in \eqn{HG} one can show that this is also the case in fixed coupling evolution  for an arbitrary configuration. For example, the quadrupole in \eqn{QconvS} assuming MV model initial conditions at $Y_0$ reads
 \begin{align}
 \label{Qsol}
 \blan \hat{Q}_{\bmx_1\bmx_2\bmx_3\bmx_4} \bran_Y
 =\,\, &\frac{\ln\big[\blan \hat{S}_{\bmx_1\bmx_2}\bran_Y
 \blan \hat{S}_{\bmx_3\bmx_4}\bran_Y
 /\blan \hat{S}_{\bmx_1\bmx_3}\bran_Y
 \blan \hat{S}_{\bmx_2\bmx_4}\bran_Y\big]}
 {\ln\big[\blan\hat{S}_{\bmx_1\bmx_2}\bran_Y
 \blan \hat{S}_{\bmx_3\bmx_4}\bran_Y
 /\blan \hat{S}_{\bmx_1\bmx_4}\bran_Y
 \blan \hat{S}_{\bmx_2\bmx_3}\bran_Y\big]}\, 
 \blan \hat{S}_{\bmx_1\bmx_2}\bran_Y
 \blan \hat{S}_{\bmx_3\bmx_4}\bran_Y
 \nn
 +\,\,
 &\frac{\ln\big[\blan \hat{S}_{\bmx_1\bmx_4}\bran_Y
 \blan \hat{S}_{\bmx_2\bmx_3}\bran_Y
 /\blan \hat{S}_{\bmx_1\bmx_3}\bran_Y
 \blan \hat{S}_{\bmx_2\bmx_4}\bran_Y\big]}
 {\ln\big[\blan\hat{S}_{\bmx_1\bmx_4}\bran_Y
 \blan \hat{S}_{\bmx_2\bmx_3}\bran_Y
 /\blan \hat{S}_{\bmx_1\bmx_2}\bran_Y
 \blan \hat{S}_{\bmx_3\bmx_4}\bran_Y\big]}\,  
 \blan \hat{S}_{\bmx_1\bmx_4}\bran_Y
 \blan \hat{S}_{\bmx_2\bmx_3}\bran_Y.
 \end{align}
At the formal level this expression was first derived in the MV model  \cite{JalilianMarian:2004da} at large-$N_c$ and later on generalized 
at finite-$N_c$ \cite{Dominguez:2011wm}. One non-trivial achievement in \cite{Iancu:2011ns,Iancu:2011nj} was to 
show when and why it remains (approximately) valid after 
quantum evolution has been taken into account. Like \eqn{QconvS} it also reduces to \eqn{Qweak} in the weak scattering limit. We note that for an arbitrary configuration in running coupling evolution one is supposed to use \eqn{QconvS}, since the arguments leading to \eqn{Qsol} do not go through (except in the case that the scale of the running coupling is taken to be 
$Q_s$). Still we shall see that, so long as $r Q_s$ is not becoming too large, one can also use the simpler version \eqn{Qsol} for practical purposes.

At finite-$N_c$ there is operator mixing and the analogue of \eqn{Qsol} involves the diagonalization of a 
matrix. For example, the quadrupole mixes with the two dipoles operator, while the 
phenomenologically interesting 6-point operator 
$\hat{Q}_{\bmx_1\bmx_2\bmx_3\bmx_4}\hat{S}_{\bmx_4\bmx_1}$, appearing in the quark-gluon double inclusive cross section, mixes with two more operators and one can 
show that the emerging cubic equation has a unique real solution. It goes without
saying, analytic expressions for high-point correlators are invaluable in order to 
reduce the numerical cost for Fourier transforming to momentum space and obtain the 
desired cross sections.

\begin{figure}
\begin{center}
\includegraphics[width=1.0\textwidth,angle=0]{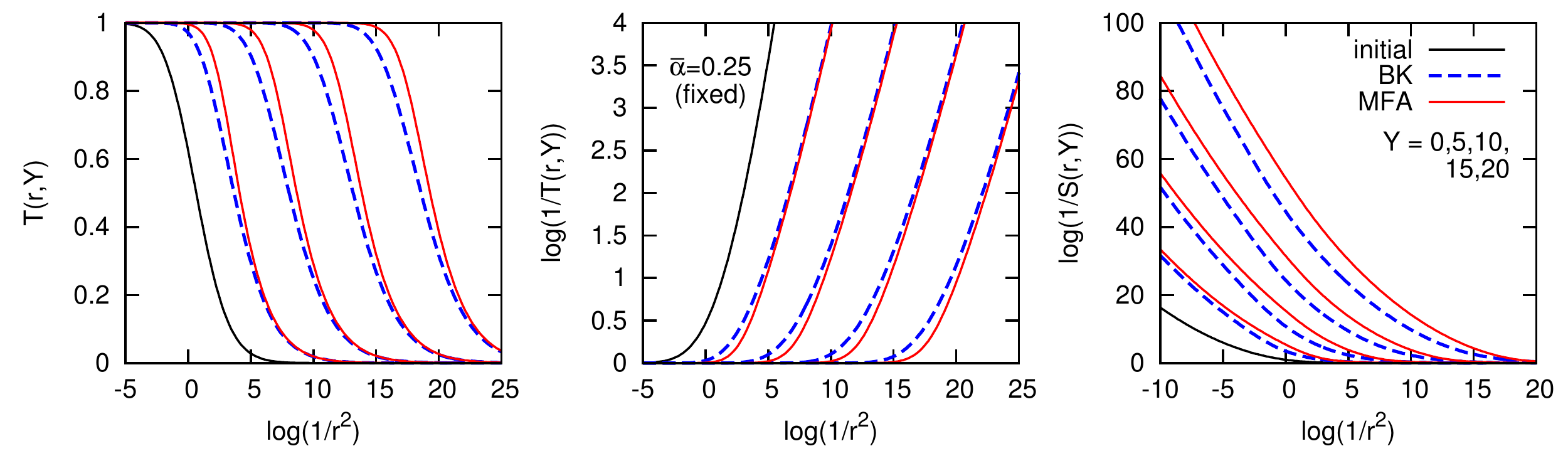}
\end{center}
\begin{center}
\includegraphics[width=1.05\textwidth,angle=0]{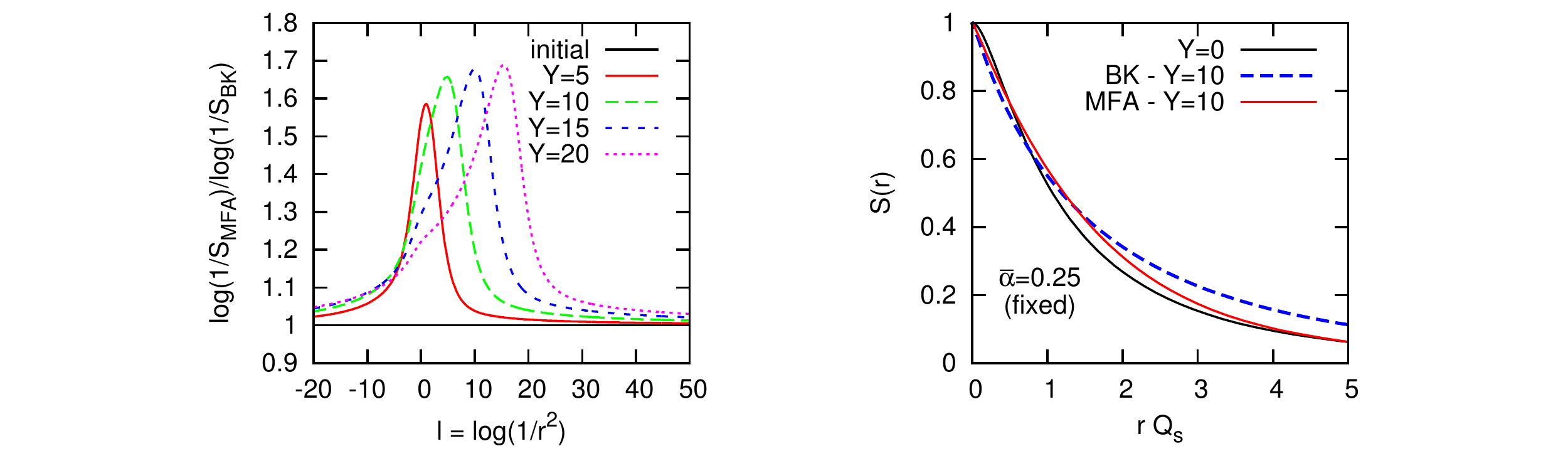}
\end{center}
\caption{Precise solution to the BK and MFA equations for fixed coupling.}
\label{fig:BKmfafc}
\end{figure}

\section{\label{sec:mfa}A mean field equation}

One way to test the accuracy of the approximation scheme is by requiring that the 
Gaussian Hamiltonian $H_{\rm G}$ in \eqn{HG} coincides with the ``averaged'' JIMWLK 
Hamiltonian $H$ in \eqn{H} \cite{Iancu:2002xk,Iancu:2011nj}. Then it is an easy and 
straightforward exercise to obtain a \emph{closed} non-linear equation for the 
dipole in the adjoint representation $A$ which reads \cite{Iancu:2011nj}
 \beq
 \label{mfa}
 \frac{\del \big\lan \hat{S}^A_{\bmx_1\bmx_2} \big\ran_Y}{\del Y} =
 \frac{\abar}{\pi}
 \int_{\bmz} \mcal{M}_{\bmx_1\bmx_2 \bmz}\,
 \frac{\big\lan \hat{S}^A_{\bmx_1\bmx_2}\big\ran_Y}
 {\big\lan 1+ \hat{S}^A_{\bmx_1\bmx_2}\big\ran_Y}
 \big\lan \hat{S}^A_{\bmx_1\bmz}
 +\hat{S}^A_{\bmz\bmx_2}
 -\hat{S}^A_{\bmx_1\bmx_2} -1
 \big\ran_Y.
 \eeq
We shall study the solution to the above in the large-$N_c$ limit, in which 
$\lan \hat{S}^A_{\bmx_1\bmx_2}\ran_Y = \lan \hat{S}_{\bmx_1\bmx_2}\ran_Y^2$ and 
then compare to that of the BK equation. A quick inspection shows that \eqn{mfa} 
reduces to the BFKL equation in the limit of weak scattering and to \eqn{BKsat}, 
like the BK equation, in the regime of strong scattering. 

This expected analytic behavior is in fact observed when we 
compare the numerical simulations of the BK equation \eqn{BK} to the simulations of the MFA equation \eqref{mfa}. 
In Fig.~\ref{fig:BKmfafc} we do see that the two 
equations, using the same initial condition, agree in the regime of weak 
scattering. When approaching the saturation regime the two solutions start to 
separate and, even though $\lan \hat{S} \ran_Y^{\rm BK}$ and  
$\lan \hat{S} \ran_Y^{\rm MFA}$ are not the same at saturation, their logarithms 
agree quite well as we can easily infer from the right plot in the upper panel and the left plot in the lower panel of Fig.~\ref{fig:BKmfafc} 
and as was advocated at the beginning.

\begin{figure}
\begin{center}
\includegraphics[width=1.0\textwidth,angle=0]{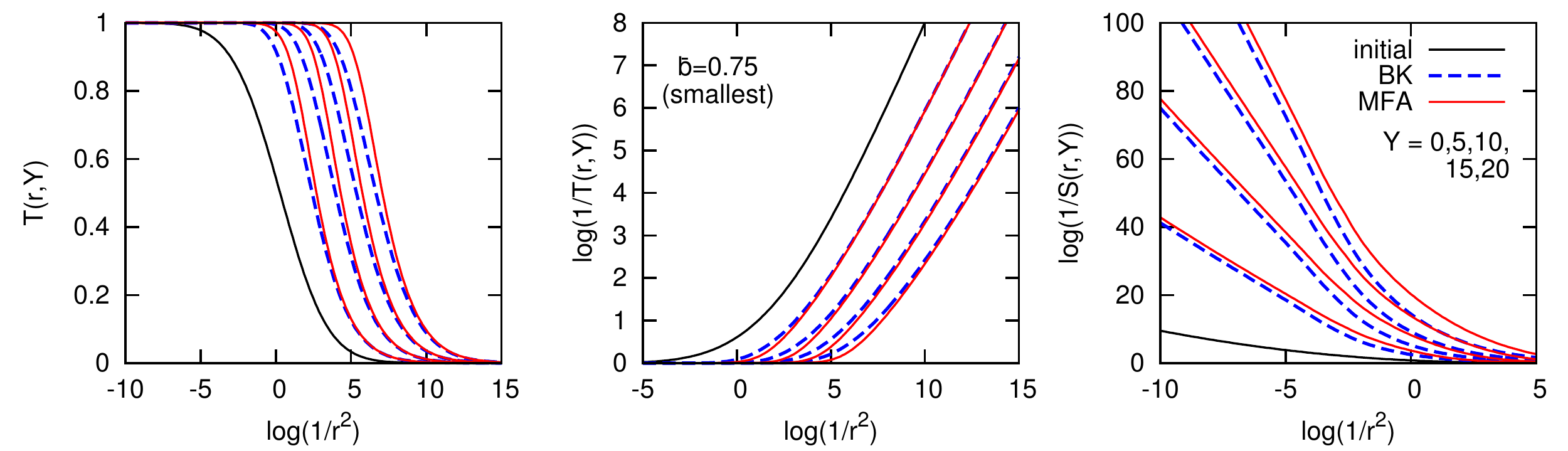}
\end{center}
\begin{center}
\includegraphics[width=1.0\textwidth,angle=0]{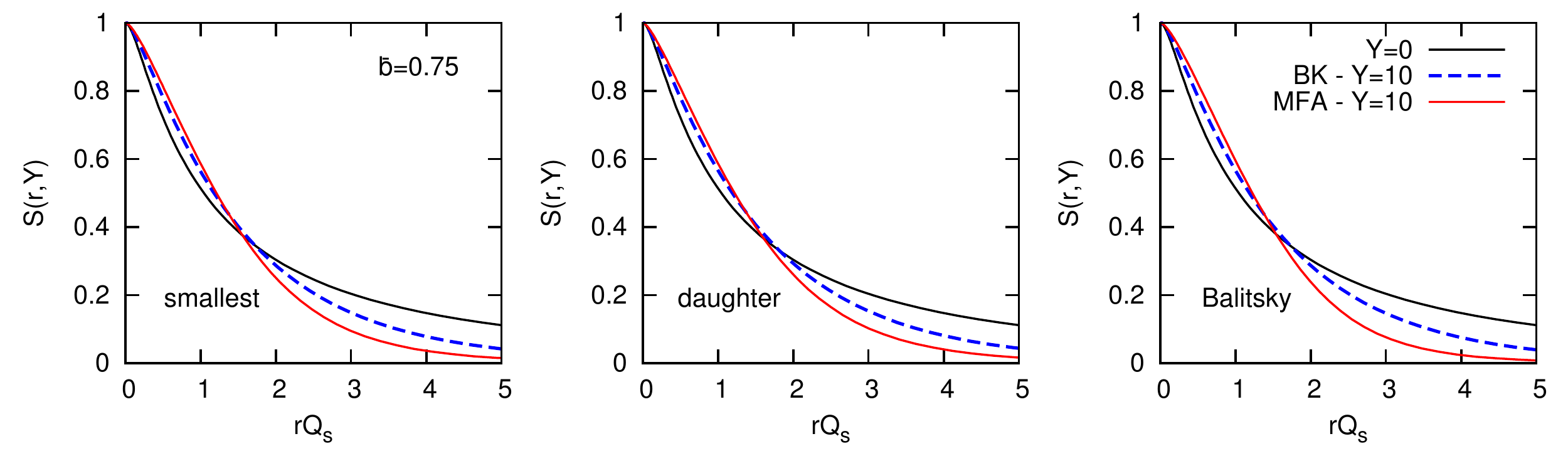}
\end{center}
\caption{Precise solution to the BK and MFA equations for running coupling. Above: smallest dipole prescription. Below: Linear scale plots for various schemes for the running of the coupling.}
\label{fig:BKmfarc}
\end{figure}

\eqn{mfa} is equally good in running coupling evolution 
so long as we adopt, as we should, the same prescription for the running of the coupling in both the BK and JIMWLK 
equations and indeed this is what is shown in
Fig.~\ref{fig:BKmfarc}. In the upper panel we show the results for the smallest dipole prescription, introduced earlier in Sect.~\ref{sec:bk}, which are similar to the fixed coupling ones in the upper panel of Fig.~\eqn{fig:BKmfafc}. Even though not shown in a logarithmic scale, to avoid a proliferation of plots, the results for the daughter dipole prescription as in \cite{Lappi:2011ju,Dumitru:2011vk} and the
Balitsky prescription as in \cite{Balitsky:2006wa}, are almost identical in shape\footnote{Here we mention the observation of a ``peculiar'', and perhaps unphysical, feature already pointed out in \cite{Berger:2011ew}: the solutions (for both BK and MFA) with the Balitsky prescription evolve slower than those with the smallest dipole one. In \cite{Balitsky:2006wa,Balitsky:2008zza} it was shown that the Balitsky prescription is equivalent to the smallest dipole one in the limits where the dipoles sizes are very different. However one can see that when the daughter dipoles are large, that is when $|\bmx_1 - \bmz| \simeq |\bmx_2 - \bmz| \gg |\bmx_1-\bmx_2|$ then one finds $\abar_{\rm Bal} \simeq \abar(|\bmx_1 - \bmx_2|)(1-4 \abar(|\bmx_1 - \bmz|))$ and therefore this convergence is very slow.}. In fact this property can be inferred from the linear plots in the lower panel of Fig.~\ref{fig:BKmfarc}.

We shall see in the next Section that, when we express higher-point correlators in 
terms of the dipole, the accuracy is not restricted to the logarithmic level.

\section{\label{sec:quad}Quadrupole configurations}

Ideally, one would like to solve the quadrupole equation in general. In principle 
this is possible since it is a closed equation (at large-$N_c$), but one 
understands that in practice it is rather complicated, because of the large number 
of transverse variables (8 in general, 6 if we assume
impact-parameter independence and 5 
if we also impose rotational invariance) on which the quadrupole depends and 
because of the non-locality in the transverse plane. Therefore we have to be more 
modest in our goals, and what we shall do is to average the quadrupole with a 
Gaussian wave-function, thus obtaining an evolution equation for the dipole. The 
solution to the latter will be compared with the solution to the BK equation, and 
such a comparison should give us a good estimate of the validity of the Gaussian 
approximation. Still, many coordinates appear in such a general equation and 
therefore we shall consider only a couple of special configurations. We 
shall immediately see that the one of them can be analytically investigated.

\begin{figure}
\begin{minipage}[b]{0.4\textwidth}
\begin{center}
\includegraphics[scale=0.6]{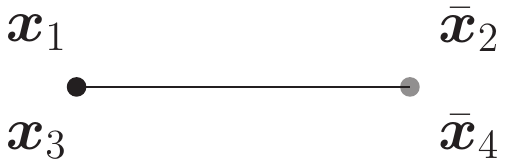}\\{\footnotesize (a)}
\end{center}
\end{minipage}
\begin{minipage}[b]{0.07\textwidth}
\begin{center}
$\xrightarrow{\hspace*{0.75cm}}$
\vspace{1.05cm}
\end{center}
\end{minipage}
\begin{minipage}[b]{0.4\textwidth}
\begin{center}
\includegraphics[scale=0.6]{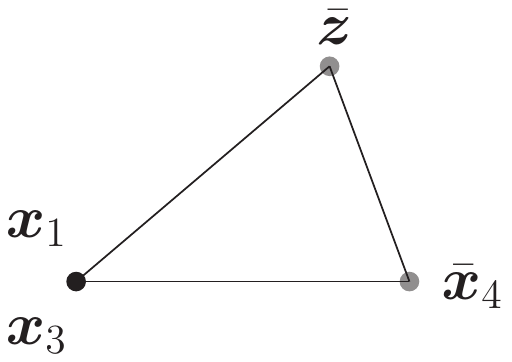}\\{\footnotesize (b)}
\end{center}
\end{minipage}
\caption{(a) The ``line'' configuration of the quadrupole for which 
$r_{12}=r_{14}=r_{23}=r_{34}$ with $r_{13}=r_{24}=0$. (b) A triangle configuration 
emerging from the evolution of the ``line''.}
\label{fig:configs1}
\end{figure}

\paragraph{The ``line'' configuration} To start, let us consider the ``line'' 
configuration of the quadrupole, as shown in Fig.~\ref{fig:configs1}.(a), where we 
put the two quarks at the same coordinate and similalry for the two antiquarks. 
This is the simplest possible one, since it is characterized by only one 
non-vanishing distance, and its evolution, according to \eqn{Qevol}, is given by
\beq\label{Qlinevol}
 \frac{\del \lan \hat{Q}_{\bmx_1\bmx_2\bmx_1\bmx_2}\ran_Y}{\del Y}=
 \frac{\abar}{\pi}
 \int_{\bmz}
 \mcal{M}_{\bmx_1\bmx_2\bmz}
 \big\lan \hat{S}_{\bmx_1\bmz}\hat{Q}_{\bmz\bmx_2\bmx_1\bmx_2}
 +\hat{S}_{\bmz\bmx_2} \hat{Q}_{\bmx_1\bmz\bmx_1\bmx_2}
 - \hat{Q}_{\bmx_1\bmx_2\bmx_1\bmx_2}
 -\hat{S}_{\bmx_1\bmx_2}^2\big\ran_Y.
 \eeq
Let us assume the large-$N_c$ limit, so that we can factorize 
$\lan\hat{S}\hat{Q}\ran_Y\to \lan\hat{S}\ran_Y \lan\hat{Q}\ran_Y$ and
$\lan\hat{S}^2\ran_Y\to\lan\hat{S}\ran_Y^2$. Then the above equation
becomes a closed equation for the quadrupole, where the dipole is
known from the solution to the BK equation. Still, as expected, this
is not a closed equation for the particular quadrupole configuration,
since a more general one appears in the real terms on the r.h.s. of
the equation. It has the shape of a triangle as shown in
Fig.~\ref{fig:configs1}.(b); only the two quarks are at the same point
with the two antiquarks being separated, or vice versa. As we have
said above, we shall assume a Gaussian average over
Eq.~\eqref{Qlinevol}, and use \eqref{Qsol}  in order to obtain an equation for the dipole. To this end we need the corresponding average for the configurations appearing in Eq.~\eqref{Qlinevol} and which are given by or obtained from (always in the large-$N_c$ limit)
 %\beq\label{Qline}
 %\lan \hat{Q}_{\bmx_1\bmx_2\bmx_1\bmx_2}\ran_Y = 
 %\lan\hat{S}_{\bmx_1\bmx_2}\ran_Y^2
 %\big[1 + \ln \lan \hat{S}_{\bmx_1\bmx_2}\ran_Y^2\big],
 %\eeq
\beq\label{Qtriangle}
 \lan \hat{Q}_{\bmx_1\bmz\bmx_1\bmx_2}\ran_Y =
 \lan\hat{S}_{\bmx_1\bmx_2}\ran_Y \lan\hat{S}_{\bmx_1\bmz}\ran_Y
 \big[1 + \ln (\lan\hat{S}_{\bmx_1\bmx_2}\ran_Y 
 \lan\hat{S}_{\bmx_1\bmz}\ran_Y/\lan\hat{S}_{\bmx_2\bmz}\ran_Y)\big],
 \eeq
and then Eq.~\eqref{Qlinevol} leads to
 \beq\label{Qlinevolave}
 \frac{\del \lan \hat{S}_{\bmx_1\bmx_2}\ran_Y}{\del Y}=
 \frac{\abar}{2 \pi}
 \int_{\bmz}
 \mcal{M}_{\bmx_1\bmx_2\bmz}
 \big[\lan \hat{S}_{\bmx_1\bmz} \ran_Y \lan\hat{S}_{\bmz\bmx_2}\ran_Y
 -\lan\hat{S}_{\bmx_1\bmx_2}\ran_Y
 \big].
 \eeq
which is nothing else than the BK equation. We would also like to stress here, 
without going through the details of the derivation, that the above is still true 
even when $N_c$ remains finite; it is a straightforward exercise to show that 
taking the Gaussian average of Eq.~\eqref{Qlinevol} we are lead to
 \beq\label{BKNc}
 \frac{\del \ln \lan \hat{S}_{\bmx_1\bmx_2}\ran_Y}{\del Y}=
 \frac{N_c^2-1}{N_c^2}\,\frac{\abar}{2 \pi}
 \int_{\bmz}
 \mcal{M}_{\bmx_1\bmx_2\bmz}
 \Bigg[\left(\frac{\lan \hat{S}_{\bmx_1\bmz} \ran_Y \lan\hat{S}_{\bmz\bmx_2}\ran_Y}
 {\lan\hat{S}_{\bmx_1\bmx_2}\ran_Y}\right)^{\frac{N_c^2}{N_c^2-1}}-1
 \Bigg],
 \eeq
which is the Gaussian average of the dipole equation at finite-$N_c$. Moreover, let 
us note that one arrives again at the above when considering the Gaussian average in 
the evolution equation of $\lan\hat{S}_{\bmx_1\bmx_2}^2\ran_Y$.

All this is a strong indication that the Gaussian approximation should be a very 
good approximation at least to the particular configurations under consideration. 
However, one should not draw the conclusion that it is the exact answer. Indeed, 
one can proceed to assume a Gaussian in the evolution equation of the triangle 
quadrupole configuration in Fig.~\ref{fig:configs1}.(b) to find that this time it 
does not reduce to the BK equation for the dipole. 

\begin{figure}
\begin{minipage}[b]{0.4\textwidth}
\begin{center}
\includegraphics[scale=0.6]{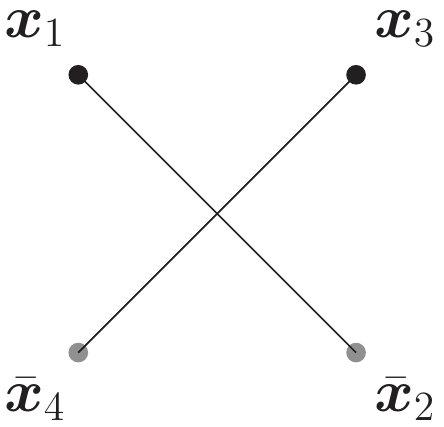}\\{\footnotesize (a)}
\end{center}
\end{minipage}
\begin{minipage}[b]{0.07\textwidth}
\begin{center}
$\xrightarrow{\hspace*{0.75cm}}$
\vspace{2cm}
\end{center}
\end{minipage}
\begin{minipage}[b]{0.4\textwidth}
\begin{center}
\includegraphics[scale=0.6]{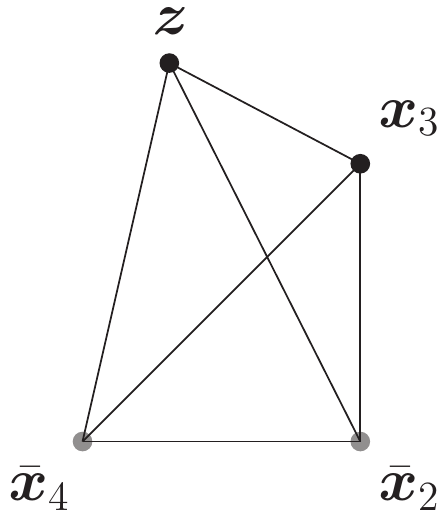}\\{\footnotesize (b)}
\end{center}
\end{minipage}
\caption{(a) The ``square'' configuration of the quadrupole for which 
$r_{13}=r_{14}$ and $r_{23}=r_{24}$. The average value of $\hat{Q}$ depends only on 
the distances depicted by straight lines. (b) A configuration emerging from the 
evolution of the ``square''.}
\label{fig:configs2}
\end{figure}

\paragraph{A ``square'' configuration} Now let us study the configuration shown in 
\ref{fig:configs2}.(a) where the four fermions are located at the corners of a 
square\footnote{Notice that this is different than the ``square'' configuration 
studied in \cite{Dumitru:2011vk}. There, the two quarks are at the edges of the one 
diagonal and similarly for the antiquarks. Here, they are at the edges of the same 
side.}. Nevertheless, we prefer to draw the diagonals, for reasons that we 
immediately explain. Clearly, a ``square'' configuration is simple enough in the sense 
that there are only two different distances (the side and the diagonal) between the 
fermions. This is one of the special class of configurations where the quadrupole 
can be written as a product of two dipoles in the Gaussian approximation 
\cite{Iancu:2011ns,Iancu:2011nj}. More precisely, for any value of $N_c$, one has
 \beq
 \lan \hat{Q}_{\bmx_1\bmx_2\bmx_3\bmx_4}\ran_Y =
 \lan\hat{S}_{\bmx_1\bmx_2}\ran_Y
 \lan\hat{S}_{\bmx_3\bmx_4}\ran_Y
 =\lan\hat{S}(R)\ran_Y^2,
 \eeq
where with $R$ we denote the length of the diagonal. The r.h.s.~of the quadrupole 
equation involves quadrupole configurations more complicated than the ``square'', 
e.g.~$\lan \hat{Q}_{\bmz\bmx_2\bmx_3\bmx_4}\ran_Y$, where $\bmz$ is located 
anywhere in the 2-dim transverse plane as shown in Fig.~\ref{fig:configs2}.(b). This also means that our test of the Gaussian approximation is probing not only the simple square configuration, but also the wider sample of configurations represented by the one in Fig.~\ref{fig:configs2}.(b).
Using the symmetries of the configuration under consideration we can regroup the terms on the r.h.s.~of the equation which simplifies to
 \begin{align}\label{Qsquare2evol}
 \hspace*{-1cm}
 \frac{\del \lan\hat{S}(R)\ran_Y}{\del Y} = 
 \frac{\abar}{2\pi}\,\frac{1}{\lan\hat{S}(R)\ran_Y}
 \int_{\bmz}
 (\mcal{M}_{\bmx_1\bmx_2\bmz} \!+\! 
 \mcal{M}_{\bmx_1\bmx_4\bmz} \!-\! 
 \mcal{M}_{\bmx_2\bmx_4\bmz})
 \big[\lan \hat{S}_{\bmx_1\bmz}\ran_Y \lan\hat{Q}_{\bmz\bmx_2\bmx_3\bmx_4}\ran_Y
 \!-\!\lan\hat{S}(R)\ran_Y^2 ].
 \end{align}
In general, the quadrupole in the Gaussian approximation and at large-$N_c$ is 
already given in \eqn{QconvS} or its simplified version \eqn{Qsol} and
we shall use the latter to rewrite \eqref{Qsquare2evol} as
  a closed equation for $\lan\hat{S}(R)\ran_Y$. Then considering 
$\lan\hat{Q}_{\bmz\bmx_2\bmx_3\bmx_4}\ran_Y$ in Eq.~\eqref{Qsquare2evol}, some 
minor cancelations occur, since 
$\lan \hat{S}_{\bmx_2\bmx_3}\ran_Y= \lan\hat{S}_{\bmx_2\bmx_4}\ran_Y$, but one 
cannot go far in simplifying the general expression in \eqn{Qsol}. As a check of our 
manipulations let us note that in the weak scattering limit
\eqn{Qsquare2evol} reduces to the BFKL equation for 
$\lan\hat{S}(R)\ran_Y = \lan\hat{S}_{\bmx_1\bmx_2}\ran_Y$, while in the 
strong scattering regime one recovers to \eqn{BKsat}. All this is natural if the 
Gaussian Hamiltonian is expected to be a good approximation scheme.

\begin{figure}
\begin{center}
\includegraphics[width=1.0\textwidth,angle=0]{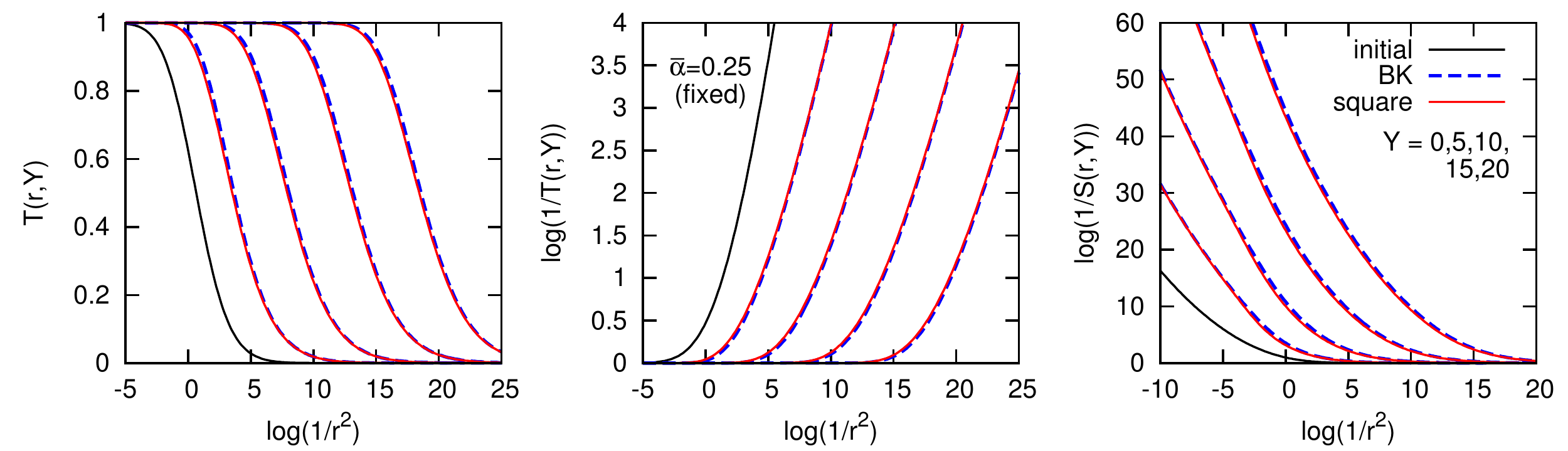}
\end{center}
\begin{center}
\includegraphics[width=1.0\textwidth,angle=0]{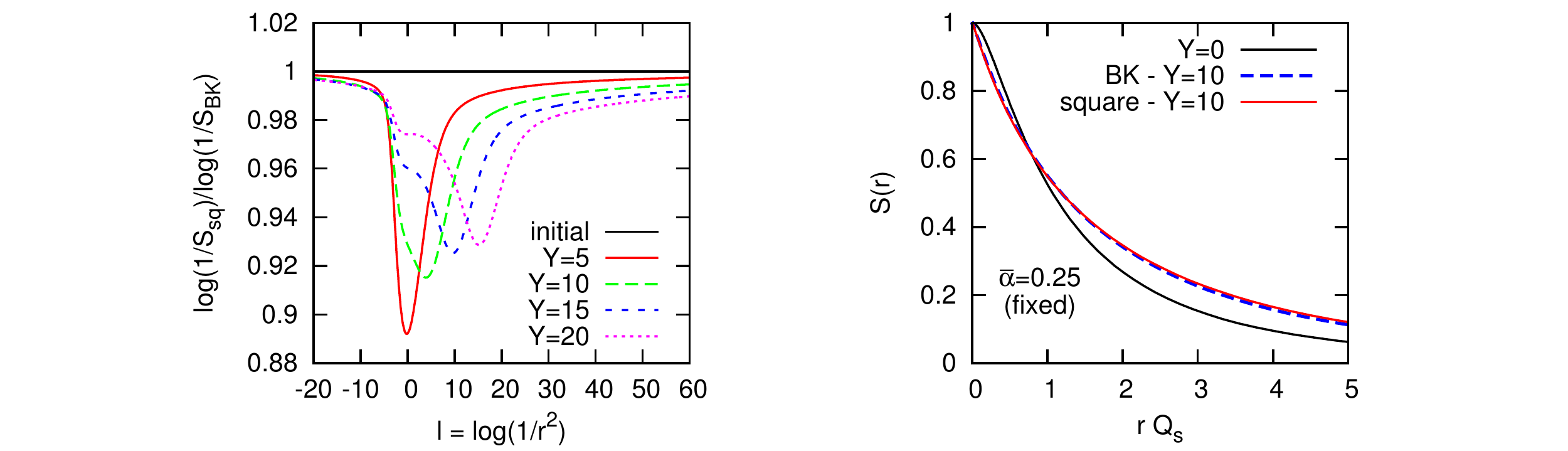}
\end{center}
\caption{Solution to the BK and the reduced quadrupole equation for the configuration in Fig.~\ref{fig:configs2} for fixed coupling.}
\label{fig:square2fc}
\end{figure}

The numerical results for fixed coupling evolution are shown in Fig.~\ref{fig:square2fc} and they should be compared to those of the MFA in Sect.~\ref{sec:mfa} shown in Fig.~\ref{fig:BKmfafc}. It is clear now that the curves arising from the solution to the BK equation and \eqn{Qsquare2evol} for the square configuration almost fall on top of each other in all kinematic regimes. This is a strong indication that the Gaussian approximation, and more precisely the extrapolation to arbitrary $Y$ of the MV model, is a quasi-exact solution to the JIMWLK equation for fixed coupling evolution.

\begin{figure}
\begin{center}
\includegraphics[width=1.0\textwidth,angle=0]{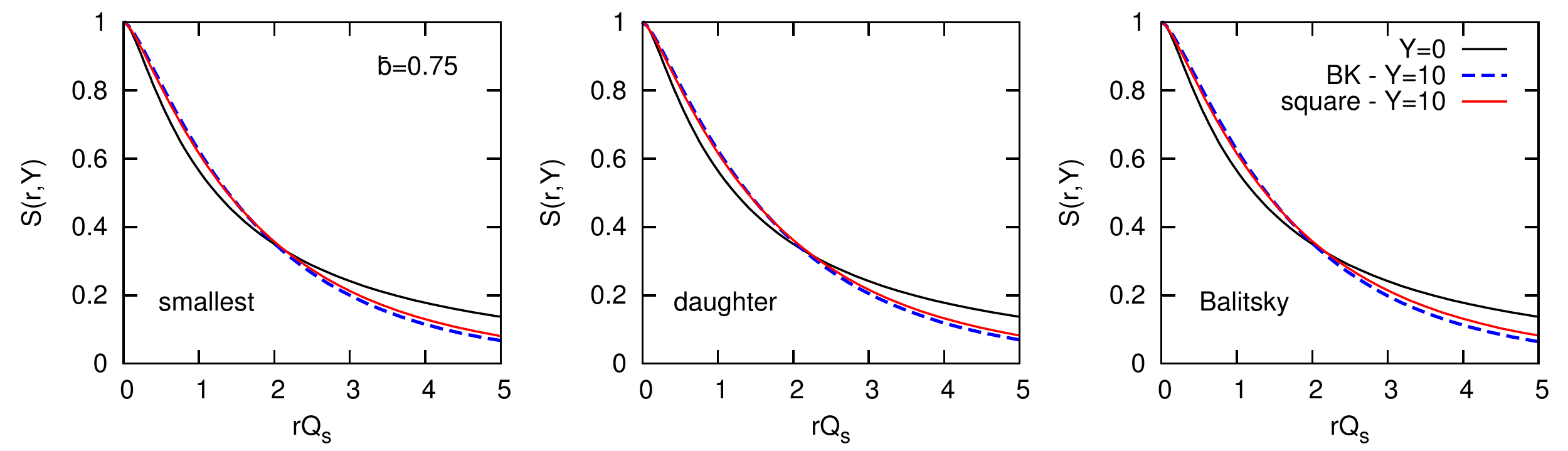}
\end{center}
\caption{Solution to the BK and the reduced quadrupole equation for our particular configuration for running coupling.}
\label{fig:square2rc}
\end{figure}

The situation is almost the same for running coupling evolution and
for all the prescriptions used, which are those adopted in
Sect.~\ref{sec:mfa}. This is exhibited in Fig.~\ref{fig:square2rc}
where one sees that in practice \eqn{Qsol} provides an excellent
approximation to the quadrupole. Some deviations start to occur when
$r Q_s$ becomes large and in fact we have seen in the numerical
solutions that these deviations grow and extend deeper in the
saturation region when rapidity the $Y$ increases, although an
agreement similar to the one observed at fixed coupling is eventually
recovered deep in the saturation region. Most likely this discrepancy
is to be attributed to the use of \eqn{Qsol} instead of the proper
expression for running coupling evolution given in \eqn{QconvS}
(cf.~the discussion above and below \eqn{Qsol}). This is also
supported by the fact that the deviations are stronger (milder) for
the Balitsky (daughter dipole) prescription  as a consequence of the
stronger (milder) variation of the coupling with the scale. Thus, it
should be interesting to check if the discrepancy goes away if one
makes use of \eqn{QconvS}, even though this looks to be mostly an
academic problem at least for this particular configuration.

\section{\label{sec:open}Correlators with open color indices}

So far we have dealt with observables which involve only traces over products of Wilson lines. However, there are quantities which require the knowledge of correlators with open color indices, like the energy density and its fluctuations immediately after the collision of two heavy ions, and we shall present here a way to calculate them. We start from the simplest possible case which is the operator $\big(V_{\bmx_1}^{\dagger} V_{\bmx_2}^{\phantom{\dagger}}\big)_{ij}$, with $i,j = 1,2, ..., N_c$. Its average value is proportional to the color structure $\delta_{ij}$, so that we must have
 \beq\label{Sopen}
 \blan\big(V_{\bmx_1}^{\dagger} 
 V_{\bmx_2}^{\phantom{\dagger}}\big)_{ij} \bran_Y= 
 \delta_{ij} \blan\hat{S}_{\bmx_1\bmx_2} \bran_Y.
 \eeq
Indeed, if we contract the above with $\delta_{ij}$ we arrive at the average of the definition of the dipole $\hat{S}_{\bmx_1\bmx_2}$ as given in \eqn{Sdipole}. This may look rather trivial for the moment but we shall see that the method works for less simple operators with more open color indices. It is also important to point out that the color structure in \eqn{Sopen} is preserved under evolution, either JIMWLK or Gaussian. At this level, one needs to be careful and use the more general form of the JIMWLK Hamiltonian  
\beq\label{Hgen}
 H = \frac{1}{8 \pi^3} \int_{\bmu\bmv\bmz}
 \mcal{K}_{\bmu\bmv\bmz}\,
 \frac{\delta}{\delta \alpha_{\bmu}^a}
 \left(1 + \wt{V}^{\dagger}_{\bmu} \wt{V}_{\bmv}^{\phantom{\dagger}}
 -\wt{V}^{\dagger}_{\bmu} \wt{V}_{\bmz}^{\phantom{\dagger}}
 -\wt{V}^{\dagger}_{\bmz} \wt{V}_{\bmv}^{\phantom{\dagger}}\right)^{ab}
 \frac{\delta}{\delta \alpha_{\bmv}^b},
 \eeq
with the kernel
 \beq\label{kernel}
 \mcal{K}_{\bmu\bmv\bmz} 
 = \frac{(\bmu-\bmz)\cdot (\bmv-\bmz)}{(\bmu-\bmz)^2(\bmv-\bmz)^2},
 \eeq
when acting on Wilson lines with open indices. Considering, as an example, only the last (the one involving $\wt{V}^{\dagger}_{\bmz} \wt{V}_{\bmv}^{\phantom{\dagger}}$) of the four terms in the parenthesis of \eqn{Hgen}, it is a straightforward exercise to show that
 \begin{align}\label{HSopen4}
 \hspace{-1.1cm}
 \blan 
 H \big(V_{\bmx_1}^{\dagger} V_{\bmx_2}^{\phantom{\dagger}}\big)_{ij}\bran_Y
 \Big|_{\rm 4th} = 
 \frac{\abar}{4 \pi}
 \int_{\bmu\bmv\bmz}
 \mcal{K}_{\bmu\bmv\bmz}\,
 (\delta_{\bmv\bmx_1} - \delta_{\bmv\bmx_2})
 \bigg[&
 \delta_{\bmu\bmx_1}\,
 \blan \hat{S}_{\bmx_1 \bmz}
 \big(V_{\bmz}^{\dagger} V_{\bmx_2}^{\phantom{\dagger}}\big)_{ij} \bran_Y
 -\delta_{\bmu\bmx_2}
 \blan\big(V_{\bmx_1}^{\dagger} V_{\bmz}^{\phantom{\dagger}}\big)_{ij}
 \hat{S}_{\bmz \bmx_2}\bran_Y
 \nn
 &-\frac{1}{N_c^2}\,
 (\delta_{\bmu\bmx_1} - \delta_{\bmu\bmx_2})
 \blan \big(V_{\bmx_1}^{\dagger} V_{\bmx_2}^{\phantom{\dagger}}\big)_{ij}\bran_Y
 \bigg].
 \end{align}
Now we employ \eqn{Sopen} in the last term in the bracket, while we furthermore assume for the first term that
 \beq\label{SOopen}
 \blan \hat{S}_{\bmx_1 \bmz}
 \big(V_{\bmz}^{\dagger} V_{\bmx_2}^{\phantom{\dagger}}\big)_{ij} \bran_Y =
 \delta_{ij}\blan \hat{S}_{\bmx_1 \bmz} \hat{S}_{\bmz \bmx_2}  \bran_Y,
 \eeq
and similarly for the second one. Making use of 
 \beq
 \int_{\bmu\bmv}
 \mcal{K}_{\bmu\bmv\bmz}\,
 (\delta_{\bmu\bmx_1} - \delta_{\bmu\bmx_2})
 (\delta_{\bmv\bmx_1} - \delta_{\bmv\bmx_2}) = 
 \mcal{M}_{\bmx_1\bmx_2\bmz},
 \eeq 
we can write
 \beq
 \blan H \big(V_{\bmx_1}^{\dagger} V_{\bmx_2}^{\phantom{\dagger}}\big)_{ij}\bran_Y
 \Big|_{\rm 4th}
 = \delta_{ij}\,\frac{\abar}{4 \pi}
 \int_{\bmz}
 \mcal{M}_{\bmx_1\bmx_2\bmz}
 \bigg\lan
 \hat{S}_{\bmx_1\bmz} \hat{S}_{\bmz\bmx_2}
 - \frac{1}{N_c^2} \hat{S}_{\bmx_1\bmx_2}
 \bigg\ran_Y
 = \delta_{ij} \blan H \hat{S}_{\bmx_1\bmx_2}\bran_Y \Big|_{\rm 4th}.
 \eeq
We can follow the same procedure for all terms of the JIMWLK Hamiltonian to finally arrive at
 \beq\label{HSopen}
 \blan H \big(V_{\bmx_1}^{\dagger} 
 V_{\bmx_2}^{\phantom{\dagger}}\big)_{ij} \bran_Y= 
 \blan H \delta_{ij} \,\hat{S}_{\bmx_1\bmx_2} \bran_Y,
 \eeq
which was our original claim. It means that if \eqn{Sopen} and equations of the form \eqref{SOopen} are valid at rapidity $Y$, then \eqn{Sopen} will hold at $Y + \Delta Y$. Subsequent action of the Hamiltonian will generate operators with more and more Wilson lines, but the only structures that can emerge are $\delta_{ij} \hat{\mcal{O}}$ (such a term in the above example is not shown in \eqn{HSopen4} but is generated by the first term of the Hamiltonian) or $(V^{\dagger}V\dots V^{\dagger}V)_{ij} \hat{\mcal{O}}$, where $\hat{\mcal{O}}$ is an operator with no open color indices. This will happen because we are interested in Wilson lines in the fundamental representation; the two functional derivatives of the Hamiltonian give rise to the color structure $(t^a)_{kl} (t^a)_{mn}$ which can be expressed in terms of Kronecker deltas, through the Fierz identity. Thus one finally arrives at the conclusion that (with $\hat{S}^{(2n)}$ given in \eqn{S2n})
 \beq\label{S2nOopen}
 \big \lan \big(V_{\bmx_1}^{\dagger} 
 V_{\bmx_2}^{\phantom{\dagger}}
 \dots
 {V}^{\dagger}_{\bmx_{2n-1}}{V}_{\bmx_{2n}}^{\phantom{\dagger}}
 \big)_{ij} \hat{\mathcal{O}}\big \ran_Y= 
 \delta_{ij} 
 \,
 \big \lan
 \hat{S}_{\bmx_1\bmx_2 ...\bmx_{2n-1}\bmx_{2n}}^{(2n)}
 \hat{\mathcal{O}}
 \big \ran_Y
 \eeq
will be true for any $\hat{\mathcal{O}}$ with no open color indices and at any value of the rapidity $Y$, since it is trivially valid for vanishing color gauge field.

Let us now go to a more interesting case by considering the operator $\big(V_{\bmx_1}^{\dagger} 
V_{\bmx_2}^{\phantom{\dagger}}\big)_{ij}
\big(V_{\bmx_3}^{\dagger} 
V_{\bmx_4}^{\phantom{\dagger}}\big)_{kl} \hat{\mcal{O}}$. Similar reasoning leads us to expect the color structure
 \beq\label{VVOopen}
 \blan \big(V_{\bmx_1}^{\dagger} 
 V_{\bmx_2}^{\phantom{\dagger}}\big)_{ij}
 \big(V_{\bmx_3}^{\dagger} 
 V_{\bmx_4}^{\phantom{\dagger}}\big)_{kl}\hat{\mcal{O}}\bran_Y 
 = \delta_{ij} \delta_{kl} \,  \blan \hat{A} \bran_Y
 + \delta_{il} \delta_{jk} \,  \blan \hat{B} \bran_Y 
 + \delta_{ik} \delta_{jl} \,  \blan \hat{C} \bran_Y,
 \eeq 
and by contracting with $\delta_{ij} \delta_{kl}$, $\delta_{il} \delta_{jk}$ and $\delta_{ik} \delta_{jl}$ we arrive at the $3 \times 3$ inhomogeneous system of linear equations
 \beq\label{ABCmat}
 \begin{bmatrix}
 N_c & 1   & 1   \\
 1   & N_c & 1   \\
 1   & 1   & N_c
 \end{bmatrix}
 \begin{bmatrix}
 \blan \hat{A} \bran_Y \\
 \blan \hat{B} \bran_Y \\
 \blan \hat{C} \bran_Y
 \end{bmatrix}
 = 
 \begin{bmatrix}
 N_c\,\blan \hat{S}_{\bmx_1\bmx_2} \hat{S}_{\bmx_3\bmx_4} \hat{\mcal{O}} \bran_Y\\
 \blan \hat{Q}_{\bmx_1\bmx_2\bmx_3\bmx_4} \hat{\mcal{O}} \bran_Y\\
 \blan \hat{R}_{\bmx_1\bmx_2\bmx_3\bmx_4} \hat{\mcal{O}} \bran_Y
 \end{bmatrix},
 \eeq
where we have defined the operator $\hat{R}_{\bmx_1\bmx_2\bmx_3\bmx_4} = (1/N_c)\big(V_{\bmx_1}^{\dagger} 
 V_{\bmx_2}^{\phantom{\dagger}}\big)_{ij} \big(V_{\bmx_3}^{\dagger} 
 V_{\bmx_4}^{\phantom{\dagger}}\big)_{ij}$. Solving \eqn{ABCmat} we obtain the gauge-invariant operators
 \begin{align}\label{Asol}
 \blan \hat{A} \bran_Y & = \frac{N_c(N_c+1) 
 \blan \hat{S}_{\bmx_1\bmx_2} 
 \hat{S}_{\bmx_3\bmx_4} \hat{\mcal{O}} \bran_Y - 
 \blan \hat{Q}_{\bmx_1\bmx_2\bmx_3\bmx_4} \hat{\mcal{O}} \bran_Y - 
 \blan \hat{R}_{\bmx_1\bmx_2\bmx_3\bmx_4}\hat{\mcal{O}} \bran_Y}{(N_c+2)(N_c-1)},
 \\\label{Bsol}
 \blan \hat{B} \bran_Y & = \frac{(N_c+1)
 \blan\hat{Q}_{\bmx_1\bmx_2\bmx_3\bmx_4} \hat{\mcal{O}} \bran_Y- 
 \blan \hat{R}_{\bmx_1\bmx_2\bmx_3\bmx_4} \hat{\mcal{O}} \bran_Y- 
 N_c\blan \hat{S}_{\bmx_1\bmx_2} 
 \hat{S}_{\bmx_3\bmx_4}\hat{\mcal{O}} \bran_Y} 
 {(N_c+2)(N_c-1)},
 \\\label{Csol}
 \blan \hat{C} \bran_Y & = \frac{(N_c+1)
 \blan \hat{R}_{\bmx_1\bmx_2\bmx_3\bmx_4} \hat{\mcal{O}} \bran_Y- 
 \blan \hat{Q}_{\bmx_1\bmx_2\bmx_3\bmx_4} \hat{\mcal{O}} \bran_Y- 
 N_c\blan \hat{S}_{\bmx_1\bmx_2} 
 \hat{S}_{\bmx_3\bmx_4}\hat{\mcal{O}} \bran_Y} 
 {(N_c+2)(N_c-1)}.
 \end{align} 
This is general and one would need to calculate separately the expectation values of the operators  $\hat{S}\hat{S}\hat{\mcal{O}} $, $\hat{Q}\hat{\mcal{O}} $ and $\hat{R}\hat{\mcal{O}}$. Of course considerable simplifications are expected to occur in the Gaussian approximation; indeed, one finds that $\hat{S}\hat{S}\hat{\mcal{O}} $, $\hat{Q}\hat{\mcal{O}} $ and $\hat{R}\hat{\mcal{O}}$ are such that 
$\blan \hat{C} \bran_Y$ in \eqn{Csol} vanishes and, thus, the color structure $\delta_{ik} \delta_{jl}$ does not appear any more in \eqn{VVOopen}. Furthermore, by eliminating $\hat{R}\hat{\mcal{O}}$, we can simplify $\blan \hat{A} \bran_Y$ and $\blan \hat{B} \bran_Y$ to finally arrive at
 \begin{align}\label{VVOopenGauss}
 \blan \big(V_{\bmx_1}^{\dagger} 
 V_{\bmx_2}^{\phantom{\dagger}}\big)_{ij}
 \big(V_{\bmx_3}^{\dagger} 
 V_{\bmx_4}^{\phantom{\dagger}}\big)_{kl}\hat{\mcal{O}}\bran_Y 
 = &\,\delta_{ij} \delta_{kl} \,
 \frac{N_c^2 \blan \hat{S}_{\bmx_1\bmx_2} 
 \hat{S}_{\bmx_3\bmx_4} \hat{\mcal{O}} \bran_Y -
 \blan \hat{Q}_{\bmx_1\bmx_2\bmx_3\bmx_4}
 \hat{\mcal{O}}\bran_Y}{N_c^2-1}
 \nn
 + &\,\delta_{il} \delta_{jk} \,
 \frac{N_c \blan  
 \hat{Q}_{\bmx_1\bmx_2\bmx_3\bmx_4}\hat{\mcal{O}} 
 \bran_Y -
 N_c \blan \hat{S}_{\bmx_1\bmx_2} 
 \hat{S}_{\bmx_3\bmx_4}
 \hat{\mcal{O}}
 \bran_Y}{N_c^2-1}.
 \end{align}
In the large-$N_c$ limit one can neglect the term involving the quadrupole in the first fraction and also set $N_c^2-1 \approx N_c^2$ in the denominators. However, one should be careful not to perform any further large-$N_c$ approximation at this stage; a measurable quantity will involve contractions over the color indices in \eqn{VVOopenGauss} and such a procedure can alter the $N_c$ counting.

\section{\label{sec:conclusion}Conclusions and perspectives}

Analytic expressions for multi-gluon correlators in the high energy limit are indispensable in order to reduce the numerical cost for obtaining the cross sections related to multi-particle production. Such expressions arise in the context of the Gaussian approximation to the JIMWLK evolution equation, and in this work we have studied the validity of such an approximation. Our results are already succinctly expressed in the last paragraph of the Introduction (Sect.~\ref{sec:intro}). In short, we have confirmed that the Gaussian approximation provides a quasi-exact solution and in particular we have shown that this is the case independent of the prescription used to set the scale in the argument of the running coupling.

Still, there remain a few things which could be addressed in a future work. It is important to explore how solid the Gaussian approximation is when we use more general initial conditions, like, for example, those in \cite{Dumitru:2011ax} and we do believe that the answer to this question is positive. Furthermore, as discussed in Sect.~\ref{sec:quad}, it would be interesting to check if the use of the more general expressions in the Gaussian approximation, like the one in \eqn{QconvS}, instead of the extrapolation of the MV model to arbitrary $Y$,  like in \eqn{Qsol}, improve the accuracy in running coupling evolution. Last, but not least, one should perhaps examine how and if the approximation applies to the case of unequal rapidity correlations.

\begin{acknowledgments}
G.S.'s work is supported by the Agence Nationale de la Recherche project
11-BS04-015-01.
Figures \ref{fig:jet}, \ref{fig:dipole}, \ref{fig:quadrupole}, \ref{fig:configs1} and \ref{fig:configs2} were created with Jaxodraw \cite{Binosi:2003yf,Binosi:2008ig}.
\end{acknowledgments}

\appendix*

\section{\label{sec:app}Numerical implementation}

In this Appendix, we sketch the main points of the numerical
techniques used to solve the evolution equations in this paper. Most
of them are common to the BK equation, that is the factorized version of \eqn{BK}, the MFA equation \eqn{mfa} and the 
equation of the square configuration \eqn{Qsquare2evol}, so we will focus on the BK equation below and
comment on the other cases after.

For the integration of the right-hand side of the equation, the first
step is to use the mirror symmetry along the bisector of the external
dipole line $x_1$, $x_2$ and limit the integration to the half-plane
containing $x_1$. In that region, we integrate in polar coordinates
around $x_1$ and easily get
 \begin{equation}
 \int_{\bmz} {\cal M}_{\bm{\bmx_1 \bmx_2 \bmz}} = 
 \int_{-\infty}^\infty \rmd \ell 
 \int_{\theta_0(\ell)}^{2\pi-\theta_0(\ell)} 
 \frac{(\bmx_1 -\bmx_2)^2}{(\bmz - \bmx_2)^2}
\end{equation}
where $\ell=\ln(1/(\bmx_1 - \bmz)^2)$ and the bounds on the $\theta$
integration --- with $\theta$ measured from the $12$ axis --- depend
on $\ell$. The radial and angular integrations are then
discretised. We typically use 16 points in $\theta$ and 1440 points in
$\ell$ with $-40\le\ell\le 140$. We have checked that for all the
plots presented in this paper, the discretisation in $\ell$ was fine
enough to reach the quoted accuracy.

Then, since deep at saturation $\lan \hat{S}(r) \ran_Y$ becomes extremely small when
rapidity increases, we decided to work instead with
$s(r,Y)=\ln(1/\lan \hat{S}(r) \ran_Y)$ and all the equations are rewritten in terms of
$s$. To achieve maximal numerical precision, terms that cancel the
logarithmic ultraviolet divergence when $\bmz$ approaches
  $\bmx_1$ are grouped.

Knowing the right-hand side of the equation, the numerical evolution in
rapidity is then handled using a standard fourth-order Runge-Kutta method.

This method applies straightforwardly to the BK and MFA equations. For the equation of the square configuration, the
bisector of the $(\bmx_1,\bmx_2)$ dipole goes through $\bmx_4$
and one may wonder about the possible ultraviolet divergence when
$\bmz$ approaches $\bmx_4$. It is easy to notice that
this appears under the form of the combination of BFKL kernels ${\cal M}_{\bmx_1 \bmx_4 \bmz}\!-\!{\cal M}_{\bmx_2 \bmx_4 \bmz}$, which is only
linearly divergent for $\bmz \to \bmx_4$, multiplied by a
combination of the dipole amplitude $S(\bmz)$ that vanishes when
$\bmz\to\bmx_4$, giving an overall smooth behaviour.

Finally, let us notice that the formulation of the evolution in terms
of $s(r,Y)=\ln(1/\lan \hat{S}(r) \ran_Y)$ also has its limitations if $Y$ becomes too
large. This is because, in order to obtain the full asymptotic
geometric scaling behaviour, one needs to describe correctly the dilute
tail over the whole geometric scaling region. Since this extends
to smaller and smaller $\lan \hat{T}(r) \ran_Y$, with $\lan \hat{T}(r) \ran_Y \approx s(r,Y)$,  when
rapidity increases, a better precision in the dilute region is
obtained by working with $\ln(1/\lan \hat{T}(r) \ran_Y)$. This can be combined with the dense region by
using a mixed variable $X(r,Y) = \max\{\ln(2/\lan \hat{S}(r) \ran_Y),\ln(2/\lan \hat{T}(r) \ran_Y)\}$. This can be
implemented using the techniques above and keeping track of the
threshold $\ell_0$ where the transition between $X=\ln(2/\lan \hat{S}(r) \ran_Y)$ and
$X=\ln(2/\lan \hat{T}(r) \ran_Y)$ occurs, {\it i.e.} where $\lan \hat{S}(\ell_0) \ran_Y=\lan \hat{T}(\ell_0) \ran_Y=1/2$. It is
this method that we have developed to verify the Levin-Tuchin formula.

%\bibliographystyle{h-physrev4}
%\bibliography{refs}

\end{document}